\documentclass{aa}
\usepackage{graphicx,amssymb,verbatim,natbib}
\bibpunct{(}{)}{;}{a}{}{,}

\newcommand\hii{\hbox{H$\rm II$~}}

\newcommand\acf{angular correlation function~}

\newcommand\wt{$w(\theta)$~}
\newcommand\se{Sect.~}

\begin{document}

 \title{The spatial clustering of radio sources in NVSS and FIRST; 
implications for galaxy clustering evolution}

\subtitle{}

   \author{R.~A. Overzier\inst{1}
          \and
          H.~J.~A. R\"ottgering
          \and
          R.~B. Rengelink
          \and R.~J. Wilman 
          }

   \offprints{R.~A. Overzier}

   \institute{Sterrewacht Leiden, PO Box 9513, 2300 RA, Leiden, The Netherlands\\
              \email{overzier@strw.leidenuniv.nl}}

   \date{Received <date> / Accepted <date>}

   \abstract{We have measured the angular correlation function, $w(\theta)$, 
of radio sources in the 1.4 GHz NVSS and FIRST radio surveys. 
Below $\sim6\arcmin$ the signal is dominated by the size distribution of classical double radio galaxies, 
an effect underestimated in previous studies. 
We model the physical size distribution of FRII radio galaxies to account for 
this excess signal in $w(\theta)$. The amplitude of the true cosmological clustering of radio sources is 
roughly constant at $A\simeq1\times10^{-3}$ for flux limits of 3--40 mJy, 
but has increased to $A\simeq7\times10^{-3}$ at 200 mJy. 
This can be explained if powerful (FRII) radio galaxies probe significantly more massive structures compared 
to radio galaxies of average power at $z\sim1$. This is consistent with 
powerful high-redshift radio galaxies generally having massive (forming) elliptical hosts in rich (proto-)cluster environments. 
For FRIIs we derive a spatial (comoving) correlation length of $r_0=14\pm3$ $h^{-1}$ Mpc. This is  
remarkably close to that measured for extremely red objects (EROs) associated with a population of old 
elliptical galaxies at $z\sim1$ by \citet{daddi01}. Based on their similar clustering properties, we propose that EROs and 
powerful radio galaxies may be the same systems seen at different evolutionary stages.   
Their $r_0$ is $\sim2\times$ higher than that of QSOs at a similar redshift, and comparable to that of bright ellipticals locally. This suggests 
that $r_0$ (comoving) of these galaxies has changed little from $z\sim1$ to $z=0$, in agreement with current $\Lambda$CDM hierarchical merging 
models for the clustering evolution of massive early-type galaxies. Alternatively, the clustering of radio galaxies can be explained by the galaxy 
conservation model. This then implies that radio galaxies of average power are the progenitors of the local field population of early-types, while 
the most powerful radio galaxies will evolve into a present-day population with $r_0$ comparable to that of local rich clusters.      

   }

\authorrunning{R.~A. Overzier et al.}
\titlerunning{The spatial clustering of radio sources}
   \maketitle


\section{Introduction}

In striking contrast with the extremely high level of isotropy observed in the temperature of the cosmic 
microwave background \citep[see e.g.][]{bernardis00}, galaxies are not distributed throughout the Universe in a random manner. 
According to the gravitational theory of instability the present structures originated from 
tiny fluctuations in the initial mass density field. This has shaped the {\it large-scale structure} of the Universe, which 
consists of vast empty regions ({\it voids}), and strings of dark and luminous matter ({\it walls}) where billions of galaxies are found. 
 
The clustering properties of galaxies can be quantified using statistical techniques, such as methods of nearest neighbours, 
counts in cells, power spectra, and correlation functions \citep[see][for an in-depth mathematical review]{peebles80}. 
In particular the two-point correlation function is a simple, but powerful tool that has become a standard for  
studying large-scale structure. The clustering of cosmological objects can be characterized by their spatial correlation 
function, which has the form $\xi(r)=(r/r_0)^{-\gamma}$ where $r_0$ is the present-day correlation length and $\gamma\simeq1.8$ 
for objects ranging from clusters to normal galaxies \citep[see][for a review]{bahcall83}. 
The local population of galaxies is a relatively unbiased tracer of the 
underlying matter distribution, with $r_0=5.4$ $h^{-1}$ Mpc derived from galaxies in the 
early CfA redshift survey by \citet{davis83}, however more recent low-redshift surveys show that 
the clustering of galaxies depends strongly on luminosity and/or morphological type. For example, 
local $L\gtrsim L_*$ ellipticals represent spatial structures that are much more strongly  
clustered with $r_0\simeq7-12$ $h^{-1}$ Mpc \citep[e.g.][]{guzzo97,willmer98,norberg02}. 
From deep, magnitude-limited redshift samples it has been found that the comoving correlation length 
of galaxies declines with redshift, roughly as expected from simple gravitational theory 
(e.g. CFRS, Le F\`evre et al. 1996; Hawaii K, Carlberg et al. 1997; CNOC2, Carlberg et al. 2000; 
CFDF, McCracken et al. 2001)\nocite{lefevre96}. In contrast to this, the clustering strength of quasars appears to vary little over 
$0\lesssim z\lesssim2.5$. \citet{croom01} found an approximately constant amplitude 
of $\sim5$ $h^{-1}$ Mpc from $\sim10,000$ quasars in the 2dF QSO Redshift Survey. 
Likewise, \citet{daddi01,daddi02} found that the (comoving) correlation length of massive elliptical 
galaxies also shows little evolution with redshift. They find $r_0=12\pm3$ $h^{-1}$ Mpc for 
a population of extremely red objects (EROs) at $z\sim1$ \citep[see also][]{mc01,roche02,firth02}, which are consistent with  
being the passively evolving progenitors of local massive ellipticals \citep[e.g.][]{dunlop96,cimatti98,cimatti02,dey99,liu00}. 
Color selection methods such as Lyman-break \citep{steidel95} and narrow-band imaging 
techniques are providing statistical samples of very high redshift galaxies, allowing us 
to study large-scale structure at even earlier epochs. Lyman-break galaxies have 
correlation lengths as high as $r_0\simeq3$ $h^{-1}$ Mpc even at $z\sim3-4$, and are thought to 
be associated with (mildly) biased star-forming galaxies \citep[e.g.][]{adelberger00,ouchi01,porciani02}. 

Studying clustering as a function of redshift and galaxy type may provide important constraints on some 
long-standing problems in cosmology concerning galaxy formation and evolution. For example, 
which of the galaxies observed at high redshift are the progenitors of local 
galaxy populations, and which of the local galaxies host the remnant black holes that 
once powered high redshift active galactic nuclei (AGN)? 
Two common views on how structures observed at high redshifts may be related to structures observed 
today are represented by (i) {\it the galaxy conservation model} \citep[e.g.][]{fry96,tegmark98} 
in which it is assumed that galaxies formed very early in a monolithic collapse \citep[e.g.][]{eggen62} and have evolved 
passively with a decreasing star formation rate since $z\sim2$, and 
(ii) {\it the hierarchical merging model}  \citep[e.g.][]{mo96} in which it is assumed that the 
most luminous galaxies formed more recently in massive dark matter haloes that have grown 
hierarchically by the merging of less massive galaxies and their haloes. 
\citet{kauffmann98} computed the evolution of the observed 
K-band luminosity function for both the monolithic case and the hierarchical case, and found 
that by a redshift of $\sim1$ these models differ greatly in the abundance of bright 
galaxies they predict. Likewise, the validity of these models can be tested by comparing  
predictions for galaxy clustering from numerical simulations or (semi-)analythic theory 
\citep[e.g.][and references therein]{kauffmann99b,moustakas02,mo02}
with the observed clustering of a population of galaxies. In the case of pure monolithic collapse galaxy clustering is dictated by 
the evolution of galaxy bias under the rules of gravitational perturbation theory, 
but without the extra non-linear effects arising from galaxy mergers. Such a scenario can be 
thought of as a baseline model for the clustering of the matter as probed by galaxies situated in average mass haloes. 
However, in the hierarchical case the evolution of galaxy bias is much more complex, since galaxies are no longer 
conserved quantitities \citep{kauffmann99b}. Comparing their observations to model predictions \citet{daddi01} find 
that such a scenario best explains the clustering evolution of massive ellipticals out to $z\sim1$. 

Radio surveys can make an important contribution to this study: 
the use of magnitude-limited surveys for finding high redshift objects is usually a cumbersome task, while 
any flux density limited sample of radio sources contains objects at redshifts of $z\sim0-5$ \citep{dp90}. 
Powerful extra-galactic radio sources, or AGN in general, result from the fuelling 
of a supermassive blackhole \citep[e.g.][]{rees84,rees90}, 
and there is evidence that the host galaxies of these high-redshift AGN are associated with some of the most 
massive structures in the early Universe 
 \citep[e.g.][]{mc88,crawford96,rott96,best98,laura99,venemans02}. Moreover, because powerful AGN were far  
more numerous at $z\sim1-2$ than today, radio surveys can be used to probe a population of 
massive galaxies in the epoch of galaxy formation. 

Despite initial concerns that any cosmological clustering of radio sources may be undetectable due 
to the relatively broad redshift distribution washing out the signal \citep[e.g.][]{webster77,griffith93},    
\citet{kooi95} detected strong clustering of bright radio sources in the 4.85 GHz 87GB survey.  
\citet{cress96} made a thorough analysis of clustering at the mJy-level. Using the 1.4 GHz FIRST survey \citep[see also][]{mag98} they 
obtained the first high-significance measurement of clustering from a deep radio sample, allowing them to investigate the separate contributions of 
both AGN and starburst galaxies \citep[but see][]{wilman02}. 
Further results on the statistics of radio source clustering have been presented by \citet{loan97} and \citet{reng98}, who based their analysis on 
the 4.85 GHz Parkes-MIT-NRAO survey and the 325 MHz WENNS survey, respectively. 
In high-resolution surveys such as FIRST, large radio sources can become resolved 
in several components, thereby spuriously contributing to the cosmological clustering signal.    
\citet{cress96} and \citet{mag98} outlined the basic steps involved in separating the signal due to this effect from the true cosmological clustering, although 
the angular size distribution of radio sources at the mJy level is still largely unconstrained. 

Since the individual redshifts of the radio sources are generally not known, one usually only measures the two-dimensional clustering 
by means of the angular correlation function, $w(\theta)$. However, the {\it redshift distribution} of the survey can be used to constrain $r_0$. 
Using this so-called Limber inversion 
technique \citep{limber53,rubin54,phil78,peebles80}, radio sources from the above surveys are typically found to have 
$r_0\approx5-15$ $h^{-1}$ Mpc. \citet{reng98} and \citet{reng99} pointed out that this broad range in $r_0$ measured can be explained by a 
scenario in which powerful radio sources have a larger $r_0$ than less powerful radio sources. This would be highly 
consistent with the mounting evidence that powerful radio galaxies are the high-redshift progenitors of local cD-galaxies residing in 
massive environments that are hence strongly clustered. Here, we will further explore the hypothesis of Rengelink et al. 
by investigating the clustering of radio sources in a number of flux-limited 
subsamples taken from the 1.4 GHz NRAO VLA Sky Survey \citep[see also][]{blake02,blake03}, 
the largest existing 1.4 GHz survey to date, containing $\sim1.8\times10^6$ radio 
sources down to a flux density limit of $\sim2.5$ mJy at $45\arcsec$ (FWHM) resolution \citep{condon98}.
We also present new results on clustering using the latest release of the FIRST survey, carefully taking into account the 
contribution of multiple-component radio sources, which we found to be severely underestimated in earlier analyses.  

The outline of this article is as follows:
in \se2 we describe our methods for measuring the angular two-point correlation function. In \se3 we describe the NVSS and FIRST radio surveys, and  
in \se4 we present measurements of the angular clustering of the sources in these surveys and construct a simple model of 
the angular size distribution of radio sources. We derive an estimate of $r_0$ as a function of flux density limit in \se5. In \se6 we compare 
our results with the results found for other populations of galaxies taken from literature, and discuss how the combined measurements relate to current 
theories on galaxy formation and evolution. The main conclusions are summarized in \se7.
 
\section{The angular correlation function} 
\label{sec:ang}

The galaxy angular two-point correlation function, $w(\theta)$, is defined as the excess 
probability, over that expected for a Poissonian distribution, of finding a galaxy 
at an angular distance $\theta$ from a given other galaxy \citep[e.g.][]{peebles80}:     
\begin{equation}
\delta P = n[1+w(\theta)]\delta\Omega,
\end{equation}
where $\delta P$ is the probablility, $n$ is the mean surface density and $\delta\Omega$ a surface
area element. The angular two-point correlation function of a given
sample of objects can be estimated as follows. For each object,
determine the angular distances to all other objects, then count the 
number of objects in each angular distance interval, denoted by $DD(\theta)$. 
As we want to calculate the {\it excess} probability of finding a galaxy at a certain distance 
from another galaxy due to clustering, we compare the
observed distribution, $DD(\theta)$, with the expected distribution of
distances, $RR(\theta)$, calculated from large artificial catalogues of
randomly placed sources. We note that several variants of $w(\theta)$-estimators exist in literature, of which 
the methods proposed by \citet{ham93} and that of \citet{landy93} \citep[see][for application of this estimator to NVSS]{blake03} are generally considered to be 
the most robust. We follow \citet{reng98} and \citet{wilman02} and use the Hamilton estimator
\begin{equation}
w(\theta)=\frac{4n_Dn_R}{(n_D-1)(n_R-1)}\frac{DD(\theta)\cdot RR(\theta)}{DR(\theta)\cdot DR(\theta)} - 1,
\end{equation}
where $n_D$ and $n_R$ are the number of sources in the data and random
catalogues, respectively, and the numerical factor $4n_D n_R / (n_D-1)(n_R-1)$ normalizes the pair counts. 
This estimator additionally makes use of the cross-correlation between
data and random catalogues, $DR(\theta)$, to minimize effects due to 
large-scale fluctuations in the mean galaxy density. We estimate $w(\theta)$ by averaging over the $w(\theta)$ computed using 
16 different random catalogues, each containing the same number of sources as the data catalogue to minimize   
the errors in $DR(\theta)$ and $RR(\theta)$ (a similar result can be  
obtained by constructing a single random catalogue that vastly exceeds the size of the data catalogue). 
Poissonian errors on the binned values of $w(\theta)$ are estimated by $\delta w(\theta)=\sqrt{[1+w(\theta)]/DD(\theta)}$.  
Alternatively, errors can be computed using the so-called bootstrap resampling method
of \citet{ling86}. In this method, the standard deviation in
$w(\theta)$ found among a large number of pseudo-random resamples 
of the original dataset is used as a measure of the error in
$w(\theta)$. However, we found that fitting a model to $w(\theta)$ (see \se\ref{sec:results}) using 
(i) Poissonian errors, and (ii) bootstrap errors gives results that are consistent within the errors of the 
fitted parameters. Therefore, given the unprecedented volumes of the radio surveys we use the first method instead of the 
relatively expensive bootstrap technique.   


\section{Survey descriptions and data selection}

\subsection{The NRAO VLA Sky Survey}

The NRAO VLA Sky Survey (NVSS) is the largest radio survey that
currently exists at 1.4 GHz. It was constructed between 1993 and 1998 
\citep{condon98}, and covers $\sim10.3$ sr of the sky north of 
$\delta=-40\degr$ ($\sim82\%$ of the sky). Fig. \ref{fig:nvss} indicates
the coverage of the NVSS. With a limiting flux density of 
$\sim 2.5$ mJy ($5\sigma_{rms}$) and an angular resolution of
$45\arcsec$ (FWHM), the NVSS contains about $1.8\times10^6$ 
sources, and is considered to be $99\%$ complete at a flux density limit of 3.4 mJy \citep{condon98}.
The NVSS is based on 217,446 snapshot observations (of mostly 23 seconds) using the VLA
in D- and DnC-configuration. These snapshots were then combined to produce
a set of $4\degr\times4\degr$ datacubes containing Stokes I, Q, and U images. 
A source catalogue was extracted by fitting the images with multiple
elliptical Gaussians. Since the angular resolution of the NVSS
($\theta\approx45\arcsec$ FWHM) is well above the median angular size of
extra-galactic radio sources ($\theta\sim10$ arcseconds), most sources 
in the catalogue are unresolved ($\gtrsim95\%$ for $3<S_{1.4}<10$ mJy). 
The main NVSS data products have been made publicly available for the use of 
the astronomical community, and can be obtained from the NRAO website\footnote{http://www.cv.nrao.edu/nvss/}.

\subsection{NVSS data selection}
\label{sec:nvsssel}

\begin{figure*}[t]
\centering
\includegraphics[width=0.75\textwidth]{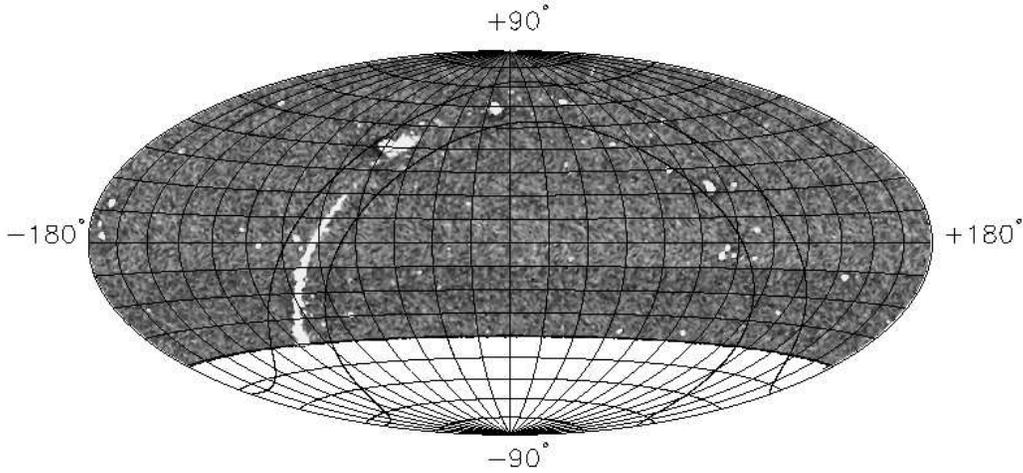}
\caption{\label{fig:nvss}Aitoff map of the NVSS source density. 
Scales run from $2\sigma$ below (black) to $2\sigma$ above the mean source density (white). 
The region of the galactic plane with $|b|<10\degr$ is indicated by solid lines. 
Besides the expected enhancement of the source density due to the large population of galactic radio sources, the NVSS catalogue 
suffers from large numbers of spurious sources around bright or extended sources (white regions), as well as an overall decrease 
in the source density below $\delta=-10\degr$ (see the greyscale change at $\delta=-10\degr$). 
See text and Table \ref{tab:excludedregions} for details.}
\end{figure*}
\begin{figure*}[ht]
\begin{minipage}[ht]{\columnwidth}
\includegraphics[width=\columnwidth]{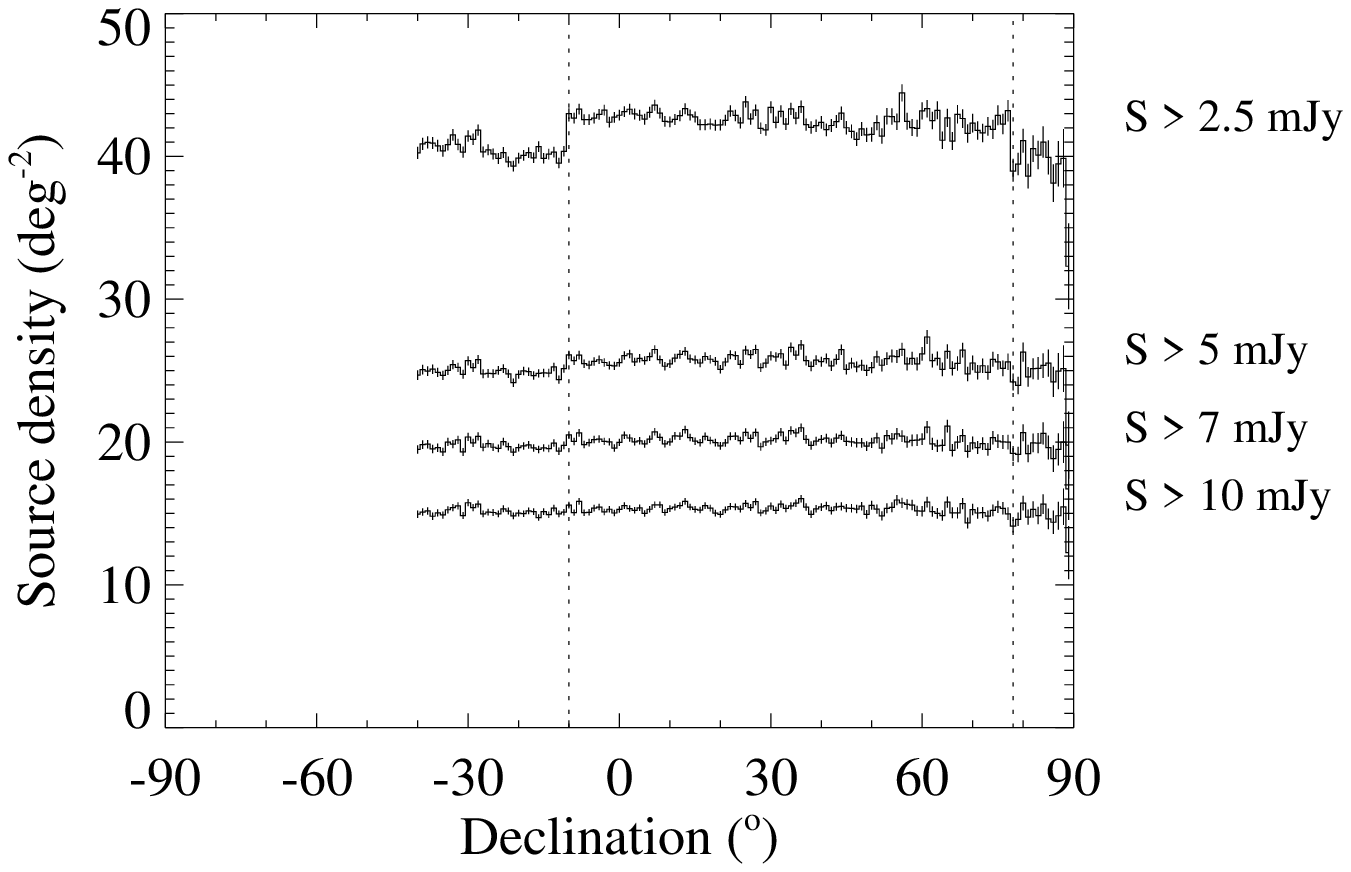}
\end{minipage}
\hfill
\begin{minipage}[ht]{\columnwidth}
\includegraphics[width=\columnwidth]{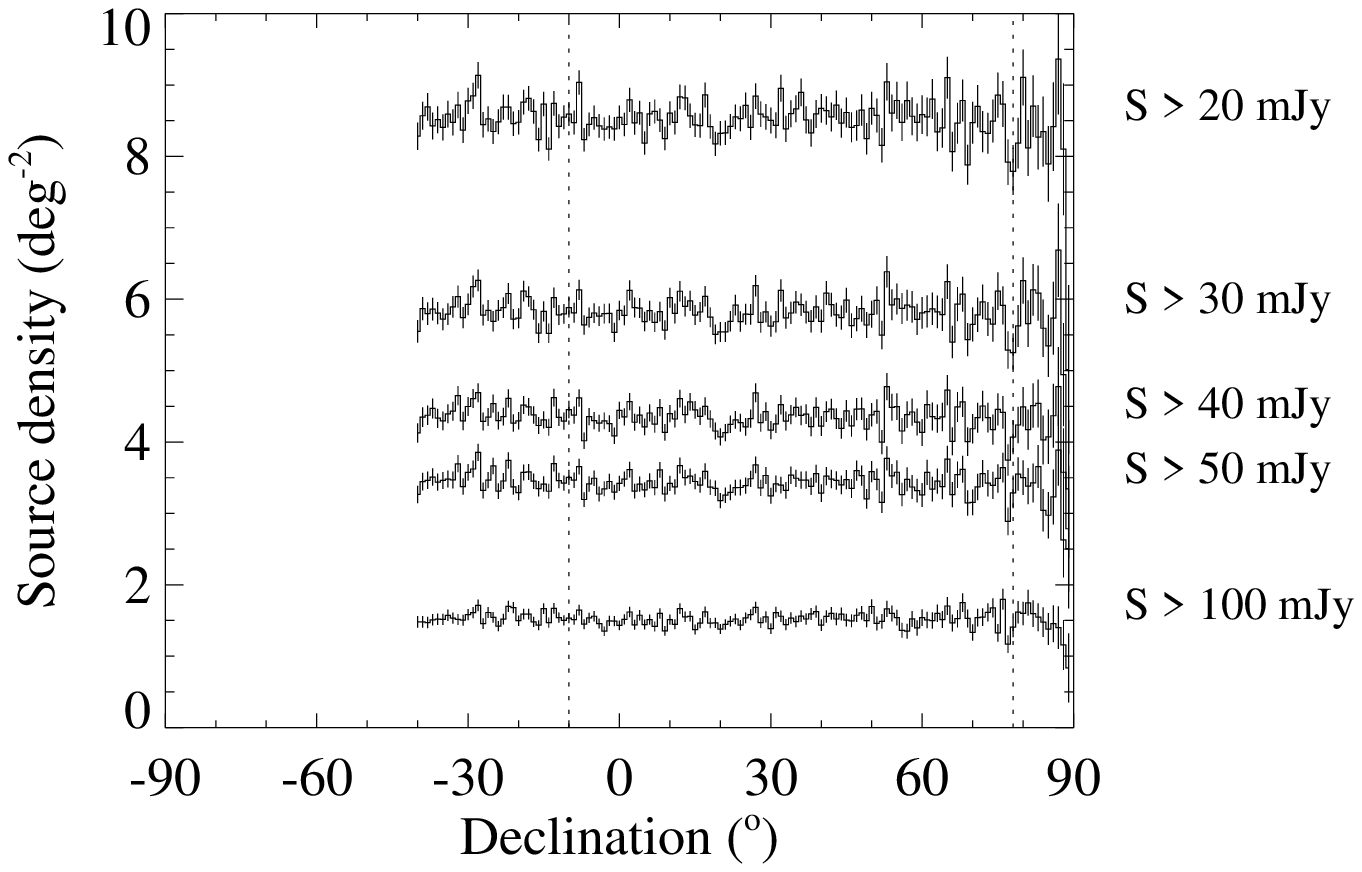}
\end{minipage}
\caption{\label{fig:decdens}The NVSS source density as a function of declination for 
various flux-limited sub-samples. Below $\sim10$ mJy beam$^{-1}$ the source density is 
non-uniform due to changes in the configuration of the VLA at $\delta=-10\degr$ and $\delta=+78\degr$ 
(dotted lines).}
\end{figure*}

To optimize our catalogue for measuring the true cosmological clustering of radio sources, we have carried out a detailed 
examination of the NVSS source catalogue to identify and correct regions that may spuriously contribute to \wt:\\ 
(i) The edge of the survey just a few arcminutes south of $\delta=-40\degr$ follows an irregular pattern with right ascension. We select 
the region $\delta\ge-40\degr$ to ensure that the boundary of the survey is straight.\\
(ii) The survey area is known to contain six hexagonal gaps due to missing snapshot 
observations that we masked from the catalogue by excluding rectangular 
regions of $2\degr\times2\degr$ fully covering each gap. The regions are listed in Table \ref{tab:excludedregions}.\\ 
(iii) We constructed a map of the NVSS source density as a function of position on the sky by applying an equal-area projection 
to the catalogue and plotting filled contours of the number of objects in $1\degr\times1\degr$ non-overlapping cells covering the survey area.  
This map is shown in Fig. \ref{fig:nvss}. The scaling of the greyscale was chosen so that underdense regions of 
$2\sigma$ below the mean density are black, and overdense regions of $2\sigma$ above the mean are white.  
Radio emission from the region of the galactic plane, as evidenced by a continuous 
chain of large white areas in Fig. \ref{fig:nvss}, is dominated by the large 
population of galactic radio sources that consists mostly of supernova remnants and \hii regions. In Fig. \ref{fig:rmsgal} we plot the rms-noise level  
as a function of galactic latitude, where the rms-noise level in each latitude bin is the average 
of the locally determined rms-noise values listed for every source entry in the NVSS catalogue. The 
rms-noise level is found to peak at $b=0\degr$ due to the overcrowding of galactic sources, but falls off to a relatively 
constant level of $\sim0.48$ mJy beam$^{-1}$ for $|b|\gtrsim10\degr$. 
We decided to exclude the region of the galactic plane that is bounded by $|b|=10\degr$, which was  
chosen so that the large overdense regions in Fig. \ref{fig:nvss} are all fully masked and the rms-noise is at a relatively constant level.\\ 
(iv) Further inspection of Fig. \ref{fig:nvss} reveals that some regions 
are associated with a significant increase in the local source density. From  
contour maps of these areas it was found that bright and/or
extended sources are sometimes accompanied by significant numbers of spurious sources due to a side-effect of the 
fitting algorithm used to extract the sources, and, in some cases, due to side-lobe contamination.  
From the catalogue we excluded rectangular regions of mostly $1\degr\times1\degr$ 
in size centered on each of these sources (larger regions of up to $2\degr\times2\degr$ were required 
in some cases). The excluded regions are listed in Table \ref{tab:excludedregions}. 
No regions of $\ge2\sigma$ underdensities were found.\\ 
\begin{figure*}[t]
\centering
\includegraphics[width=0.75\textwidth]{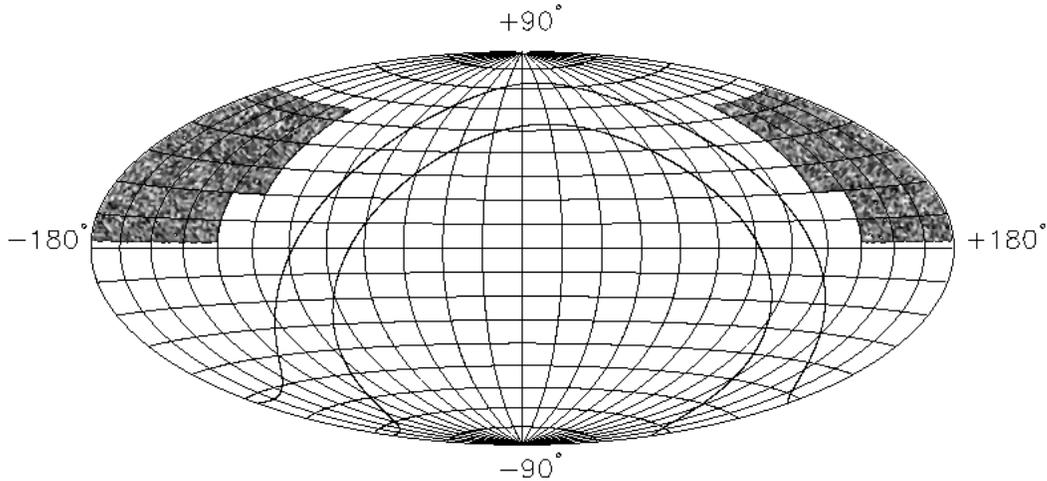}
\caption{\label{fig:firstcov}Aitoff map of the FIRST source density. 
Scales run from $2\sigma$ below (black) to $2\sigma$ above (white) the mean source density. 
The region of the galactic plane with $|b|<10\degr$ is indicated by solid lines.} 
\end{figure*}
\begin{figure}[t]
\centering
\includegraphics[width=\columnwidth]{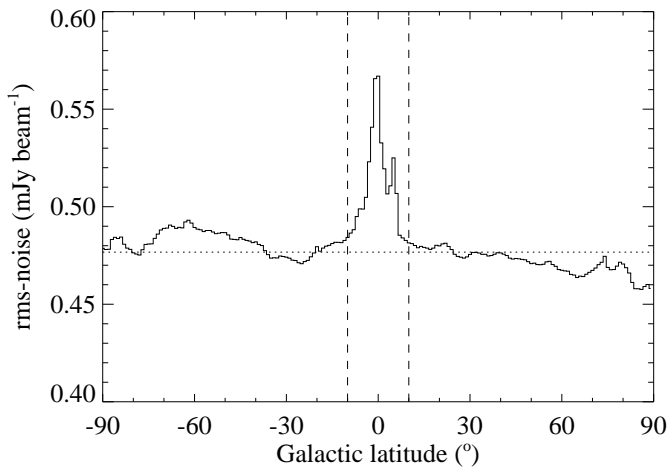}
\caption{\label{fig:rmsgal}The rms-noise level as a function of galactic latitude. The average rms-noise level of the survey is 
$\sim0.48$ mJy beam$^{-1}$. Dotted lines enclose the region $|b|<10\degr$.}
\end{figure}
(v) Most of the NVSS observations were conducted using the VLA in
D-configuration, but the regions $\delta\le-10\degr$ and $\delta\ge+78\degr$ were observed using the hybrid
DnC-configuration to counterbalance projection effects which result from foreshortening of the north-south $uv$-coverage range. 
Fig. \ref{fig:decdens} shows the NVSS source density as a function of
declination for various flux-limited sub-samples. Below the flux density limit of 10 mJy, the use of the DnC-configuration has 
caused a significant decrease in sensitivity leading to a drop in the source density of $\gtrsim10$ percent 
(see also Fig. \ref{fig:nvss}). As this will inevitably cause spurious signal in the angular two-point correlation function, 
we selected only the regions observed in D-configuration for measuring $w(\theta)$ below flux density limits of 10 mJy.\\ 

Table \ref{tab:samples} lists the final regions and the number of sources in them for various flux density limited subsamples. 
\begin{table*}[t]
\caption[]{\label{tab:excludedregions}Regions of the NVSS catalogue that were masked because of missing snapshot observations and 
overdense regions associated with bright or extended sources. Overdense regions at $|b|<10\degr$ are not listed here since 
we excluded this area from the catalogue as a whole.}
\begin{center}
\begin{tabular}[t]{ccl}
\hline
\hline
\multicolumn{1}{c}{RA (J2000)} & \multicolumn{1}{c}{DEC (J2000)} & \multicolumn{1}{c}{Remark} \\
\hline
$15^h38^m00^s$ -- $15^h43^m00^s$ & $-05\degr00\arcmin00\arcsec$ -- $-06\degr00\arcmin00\arcsec$ & Missing snapshot\\
$09^h54^m00^s$ -- $10^h00^m00^s$ & $-11\degr45\arcmin00\arcsec$ -- $-12\degr45\arcmin00\arcsec$ & Missing snapshot\\
$09^h54^m00^s$ -- $10^h00^m00^s$ & $-25\degr00\arcmin00\arcsec$ -- $-26\degr00\arcmin00\arcsec$ & Missing snapshot\\
$04^h25^m00^s$ -- $04^h30^m00^s$ & $-37\degr45\arcmin00\arcsec$ -- $-38\degr45\arcmin00\arcsec$ & Missing snapshot\\
$18^h17^m00^s$ -- $18^h22^m00^s$ & $-16\degr00\arcmin00\arcsec$ -- $-17\degr00\arcmin00\arcsec$ & Missing snapshot\\
$18^h02^m00^s$ -- $18^h07^m00^s$ & $-23\degr30\arcmin00\arcsec$ -- $-24\degr30\arcmin00\arcsec$ & Missing snapshot\\
$01^h34^m00^s$ -- $01^h42^m00^s$ & $+32\degr30\arcmin00\arcsec$ -- $+33\degr50\arcmin00\arcsec$ & 3C 48\\
$03^h16^m48^s$ -- $03^h21^m48^s$ & $+40\degr00\arcmin42\arcsec$ -- $+43\degr00\arcmin42\arcsec$ & Perseus A\\
$03^h17^m00^s$ -- $03^h27^m00^s$ & $-36\degr20\arcmin00\arcsec$ -- $-38\degr20\arcmin00\arcsec$ & Fornax A\\
$04^h35^m05^s$ -- $04^h39^m05^s$ & $+29\degr10\arcmin12\arcsec$ -- $+30\degr10\arcmin12\arcsec$ & 3C 123\\	  
$05^h18^m00^s$ -- $05^h26^m00^s$ & $-35\degr40\arcmin00\arcsec$ -- $-37\degr20\arcmin00\arcsec$ & PKS 0521--36\\	  
$05^h31^m17^s$ -- $05^h39^m17^s$ & $-06\degr23\arcmin00\arcsec$ -- $-04\degr23\arcmin00\arcsec$ & M42\\	  
$05^h38^m00^s$ -- $05^h46^m00^s$ & $-01\degr20\arcmin00\arcsec$ -- $-02\degr40\arcmin00\arcsec$ & 3C 147.1\\ 
$05^h40^m36^s$ -- $05^h44^m36^s$ & $+49\degr21\arcmin07\arcsec$ -- $+50\degr21\arcmin07\arcsec$ & 3C 147\\	  
$05^h52^m00^s$ -- $05^h56^m00^s$ & $-04\degr20\arcmin00\arcsec$ -- $-05\degr40\arcmin00\arcsec$ & TXS 0549--051\\
$07^h05^m00^s$ -- $07^h25^m00^s$ & $+74\degr20\arcmin00\arcsec$ -- $+75\degr20\arcmin00\arcsec$ & 3C 173.1\\
$09^h15^m05^s$ -- $09^h21^m05^s$ & $-12\degr50\arcmin24\arcsec$ -- $-11\degr20\arcmin24\arcsec$ & Hydra A\\	  
$12^h16^m00^s$ -- $12^h23^m00^s$ & $+05\degr00\arcmin00\arcsec$ -- $+06\degr30\arcmin00\arcsec$ & NGC 4261\\
$12^h26^m07^s$ -- $12^h32^m07^s$ & $+01\degr18\arcmin00\arcsec$ -- $+02\degr48\arcmin00\arcsec$ & 3C 273\\  
$12^h26^m50^s$ -- $12^h34^m50^s$ & $+11\degr23\arcmin24\arcsec$ -- $+13\degr23\arcmin24\arcsec$ & M87\\	  
$13^h11^m00^s$ -- $13^h15^m00^s$ & $-22\degr30\arcmin00\arcsec$ -- $-21\degr30\arcmin00\arcsec$ & MRC 1309--216\\
$13^h21^m00^s$ -- $13^h27^m00^s$ & $+31\degr20\arcmin00\arcsec$ -- $+32\degr20\arcmin00\arcsec$ & NGC 5127\\
$14^h07^m00^s$ -- $14^h15^m00^s$ & $+51\degr40\arcmin00\arcsec$ -- $+52\degr40\arcmin00\arcsec$ & 3C 295\\  
$16^h49^m11^s$ -- $16^h53^m11^s$ & $+04\degr29\arcmin24\arcsec$ -- $+05\degr29\arcmin24\arcsec$ & Hercules A\\
$17^h18^m00^s$ -- $17^h24^m00^s$ & $-01\degr40\arcmin00\arcsec$ -- $-00\degr20\arcmin00\arcsec$ & 3C 353\\
$18^h26^m00^s$ -- $18^h34^m00^s$ & $+48\degr00\arcmin00\arcsec$ -- $+49\degr40\arcmin00\arcsec$ & 3C 380\\
$19^h22^m00^s$ -- $19^h26^m00^s$ & $-28\degr45\arcmin00\arcsec$ -- $-29\degr45\arcmin00\arcsec$ & TXS 1921--293\\	  
\hline
\end{tabular}
\end{center}
\end{table*}

\subsection{The FIRST Survey}

\begin{figure}[t]
\centering
\includegraphics[width=\columnwidth]{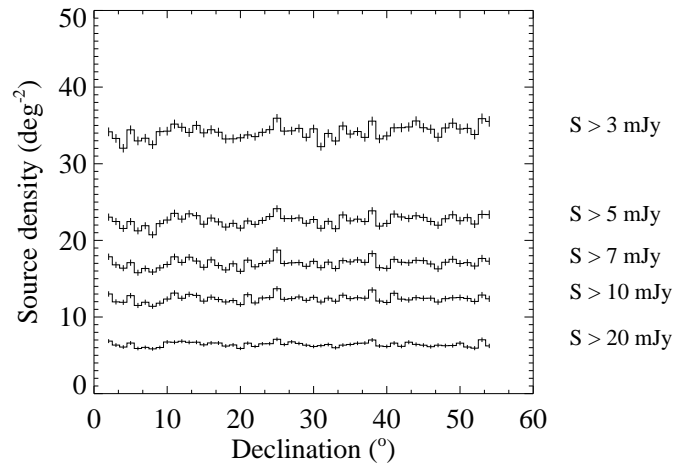}
\caption{\label{fig:decdensfirst}The FIRST source density as a function of declination for 
various limiting flux densities.}
\end{figure}
The FIRST (Faint Images of the Radio Sky at Twenty centimeters) survey
\citep{becker95} is another 1.4 GHz VLA survey, which
was started in 1993 and is still under construction. Using the VLA in
B-configuration it will ultimately cover $\sim10,000$ square degrees of the
northern Galactic cap, matching the survey area of the Sloan Digital
Sky Survey. Given the large coverage of FIRST, its sensitivity is
unprecedented: with a limiting flux density of $\sim1$ mJy
($5\sigma_{rms}$) and an angular resolution of $5\farcs4$ (FWHM) the
catalogue contains about 100 sources per square degree with a 
completeness level of $\sim95\%$ at 2 mJy \citep{becker95}.  

We have obtained the publicly available 2001 October 15 version of the
source catalogue\footnote{http://sundog.stsci.edu/}, which
has been derived from the 1993 through 2001 observations, and covers
about 8,565 square degrees of the sky. About 4\% of the 771,076 sources in the
catalogue are flagged as possible side-lobes, which we exclude 
from the catalogue. We set the lower flux density limit of the catalogue to 3 mJy, 
the limiting flux density of the NVSS survey.  
Finally, we select the regions $+2\degr\le\delta\le+20\degr$ and
$9^h\le\alpha\le16^h$, $+20\degr\le\delta\le+55\degr$ and
$8^h\le\alpha\le17^h$ from the catalogue, by requiring a relatively uniform source
density and a simple geometric form. This area covers $\sim5,538$
square degrees and contains 188,885 sources. As for the NVSS, we construct 
a map of the FIRST surface density (Fig. \ref{fig:firstcov}). and plot
the source density as a function of declination (Fig. \ref {fig:decdensfirst}). 
For the selected region we found no suspicious features in the catalogue. 
The number of sources in various FIRST subsamples are listed in Table \ref{tab:samples}. 


\section{The angular clustering of radio sources}
\label{sec:results}

\subsection{The angular correlation function of $S>10$ mJy NVSS sources} 
\label{sec:result10}

Following the procedures described in \se\ref{sec:ang} we compute \wt 
for the $S>10$ mJy NVSS subsample. Distances between data and/or random 
positions are initially measured in bins of $0\farcm5$, and
rebinned in bins of constant logarithmic spacing to analyse the data. 
We fit the data using a weighted $\chi^2$-minimization routine, 
and we determine the $1\sigma$ errors from the covariance matrix.
\begin{table*}[t]
\caption[]{\label{tab:samples}NVSS and FIRST subsamples.}
\begin{center}
\begin{tabular}[t]{llrllr}
\hline
\hline
\multicolumn{3}{c}{NVSS}&\multicolumn{3}{c}{FIRST}\\
\hline
$S_{low}$ & Region & Sources & $S_{low}$ & Region & Sources\\
\hline
3 mJy &  $+10\degr\le b\le+45\degr$, $-5\degr\le\delta\le+70\degr$ &
210,530 & 
3 mJy &$+2\degr\le\delta\le+20\degr$ and $9^h\le\alpha\le16^h$  & 188,885 \\ 
  &    &    & &  $+20\degr\le\delta\le+55\degr$ and
$8^h\le\alpha\le17^h$           &   \\
5 mJy &  $|b|\ge 10\degr$, $-10\degr\le \delta \le +78\degr$ & 507,608 & 5 mJy &\multicolumn{1}{c}{``\quad``}&124,974\\
7 mJy & $|b|\ge10\degr$, $-5\degr\le\delta\le+70\degr$  & 351,079 & 7 mJy &\multicolumn{1}{c}{``\quad``}& 94,099\\
10 mJy & $|b|\ge 10\degr$ & 433,951 & 10 mJy&\multicolumn{1}{c}{``\quad``}& 68,560\\
20 mJy &$|b|\ge 10\degr$  & 242,599 &&\\
30 mJy &$|b|\ge 10\degr$  & 165,459&&&\\
40 mJy &$|b|\ge 10\degr$  & 123,769&&&\\
50 mJy &$|b|\ge 10\degr$  & 97,753&&&\\
60 mJy &$|b|\ge 10\degr$  & 79,738&&&\\
80 mJy &$|b|\ge 10\degr$  & 56,903&&&\\
100 mJy & $|b|\ge 10\degr$ & 43,294&&&\\
200 mJy & $|b|\ge 10\degr$ & 17,015&&&\\
\hline
\end{tabular}
\end{center}
\end{table*}
\begin{figure}[ht]
\centering
\includegraphics[width=\columnwidth]{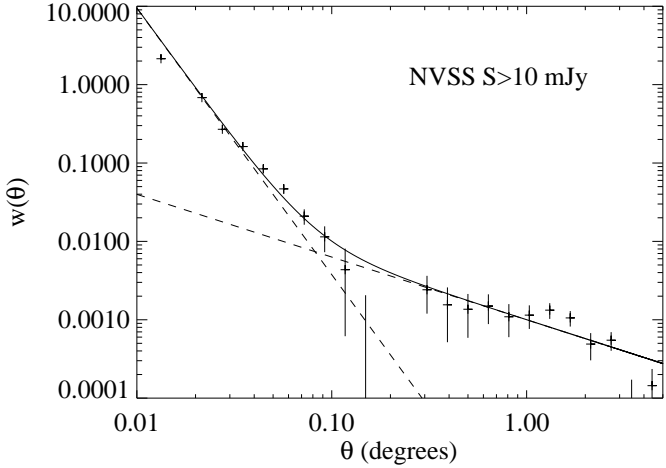}
\caption{\label{fig:10log}The angular two-point correlation function
  of $S>10$ mJy NVSS sources. The power-law fits described in the text
  are indicated.}
\end{figure}

The results are shown in Fig. \ref{fig:10log}. 
We find that two power-laws are needed to describe the full range of our measurements. 
Fitting the data with a power-law angular correlation function 
$w(\theta)=A\theta^{1-\gamma}$ \citep[e.g.][]{peebles80} at angular scales of 
$\theta\lesssim6\arcmin$ gives a slope of $\gamma=4.4\pm0.2$, while 
at $\theta\gtrsim6\arcmin$ we find a slope of $\gamma=1.7\pm0.1$. The latter value is consistent with 
the slope of the empirical power-law of $\gamma\simeq1.8$ found for the cosmological clustering of objects
ranging from normal galaxies to clusters \citep[see][for a review]{bahcall83}. However, at small 
angular scales the power-law is much steeper, presumably caused by the enhancement of $DD(\theta)$ 
 due to the decomposition of large radio galaxies into their separate 
radio components (see \se\ref{sec:first} \& \se\ref{sec:model}; see also Blake \& Wall 2002a). 
If we fit the data simultaneously with a double power-law correlation 
function of the form $w(\theta)=B\theta^{1-\gamma_B}+A\theta^{1-\gamma_A}$ with fixed slopes of 
$\gamma_B=4.4$ and $\gamma_A=1.8$, we find amplitudes of 
$B=(1.5\pm0.2)\times10^{-6}$ and $A=(1.0\pm0.2)\times10^{-3}$. The double power-law fit is indicated in Fig. \ref{fig:10log}. 

\begin{figure}[t]
\centering
\includegraphics[width=\columnwidth]{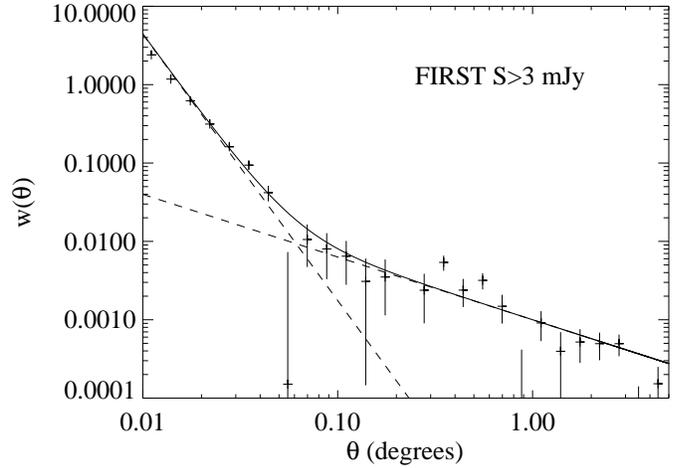}
\caption{\label{fig:first}The angular two-point correlation function
  of $S>3$ mJy FIRST sources. The power-law fits described in the text
  are indicated.}
\end{figure}

\subsection{The effect of multiple component radio sources and the \acf of FIRST}
\label{sec:first}

Although the median angular size of radio sources is $\sim10\arcsec$ \citep[e.g.][]{condon98}, 
radio sources can have sizes of up to several arcminutes.
At angular scales comparable to the size of these large radio
galaxies, the true cosmological \wt can become confused or even dominated by resolving
these galaxies into their various radio components, 
such as lobes, hot spots and cores. The angular
scale at which the size distribution of radio galaxies begins to
dominate \wt is indicated by the clear break around $6\arcmin$. 
Earlier studies attempted to correct \wt for the contribution of multi-component
radio sources by means of component combining algorithms. For example,
\citet{cress96} calculated the \acf for the FIRST survey considering all
sources within $1\farcm2$ of each other as a single source. The
analysis of the FIRST data was repeated by \citet{mag98}, who 
removed double sources using an algorithm based on the 
$\theta\propto\sqrt{S}$ relation of \citet{oort87} and flux ratio statistics 
of the components of genuine doubles. They found
values of $\gamma=2.5\pm0.1$, and $A=(1.0\pm0.1)\times10^{-3}$ for
flux density limits between 3 and 10 mJy. Comparing their results
 to our measurement for the NVSS presented in Fig. \ref{fig:10log}, we conclude that despite the efforts of these authors it is likely 
that a residual contribution from large radio galaxies remained. 
Fitting the data over the whole range of $\theta$ with a {\it single} power-law explains the apparently high
value of $\gamma\simeq2.5$ reported for the clustering of FIRST radio sources.   
\begin{figure}[t]
\centering
\includegraphics[width=\columnwidth]{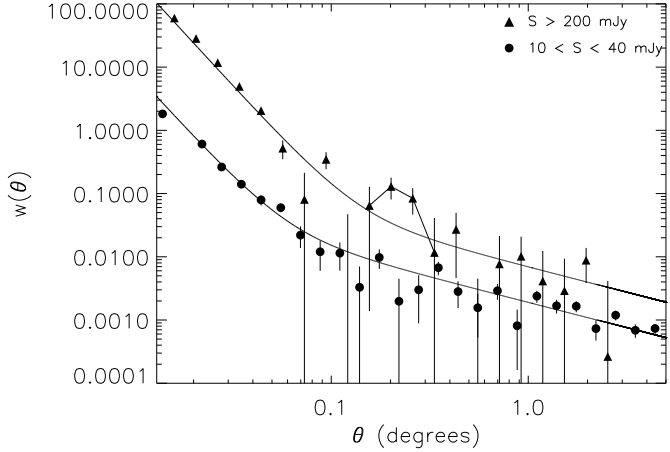}
\caption{\label{fig:comp_nvss}Angular correlation functions for the flux density intervals $10<S<40$ mJy and $S>200$ mJy. 
The power-law fits to the data described in the text are overplotted. Because of an unexplained 'bump' in the $S>200$ mJy 
signal at $0.1\lesssim\theta\lesssim0.3$ (connected points), the small- and large-scale correlation functions were fitted separately 
over the ranges $\theta\le0.1$ and $\theta\ge0.3$, respectively.}
\end{figure}

Here, we present new measurements from the FIRST survey. Our reasons for repeating the work of \citet{cress96} and \citet{mag98} are threefold. 
Firstly, the FIRST catalogue has almost doubled in size, enabling a better statistical measure of $w(\theta)$. 
Secondly, the clear break found in the 
\acf of the NVSS enabled us to isolate the signal due to true clustering from
the signal due to the size distribution of radio galaxies. A similar analysis can be applied 
to the FIRST data. Thirdly, we found large-scale gradients in 
the NVSS source density below a flux density limit of 10 mJy (see \se\ref{sec:nvsssel}). 
The FIRST data can be used to verify and complement the results from the NVSS for 3--10 mJy.         

In Fig. \ref{fig:first} we present our measurements for the \acf from the $S>3$ mJy 
FIRST subsample. As for the NVSS, we see a clear break in \wt due to the presence of multi-component radio sources. 
Fitting the measurements with our double power-law model yields $\gamma_B=4.1\pm0.2$ and $B=(2.7\pm0.3)\times10^{-6}$, 
and $\gamma_A=1.9\pm0.2$ and $A=(1.0\pm0.3)\times10^{-3}$.
Note that the break in $w(\theta)$ in this sample occurs at $\theta\sim4\arcmin$ compared to $\theta\sim6\arcmin$  
 for $S>10$ mJy in NVSS (see Fig. \ref{fig:10log}). \citet{blake02} show that this is due to a $1/\sigma$ dependency 
($\sigma$ being the surface density of radio sources) of the amplitude of \wt at small angular scales, simply because 
the weight of pair-counts due to large radio galaxies increases as the surface density decreases (see their equation 4). 

We conclude that the cosmological \wt of $S>10$ mJy NVSS sources and $S>3$ mJy FIRST sources, 
as determined by our analysis, are consistent with having the canonical clustering power-law slope 
of $\gamma\simeq1.8$, and an amplitude of $A\simeq1\times10^{-3}$. 
\begin{figure}[t]
\centering
\includegraphics[width=\columnwidth]{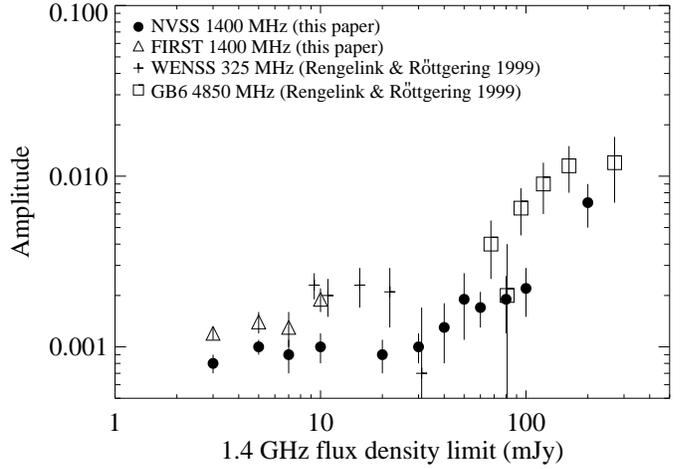}
\caption{\label{fig:wamps}The amplitude of the cosmological \acf
  ($\gamma=1.8$) of NVSS and FIRST as a function of 1.4 GHz flux density
  limit. For comparison, we have indicated the results for the WENSS and GB6 surveys from \citet{reng98} and \citet{reng99}.}
\end{figure}

\subsection{$w(\theta)$ as a function of flux density limit}
\label{sec:w_func_flux}

To investigate angular clustering as a function of flux
density limit, we calculate $w(\theta)$ for all NVSS and FIRST subsamples
listed in Table \ref{tab:samples}. We obtain the amplitudes of \wt 
by fitting the data with the double power-law model $w(\theta)=B\theta^{1-\gamma_B}+A\theta^{1-\gamma_A}$, fixing the slopes at
$\gamma_B=4.4$ and $\gamma_A=1.8$. However, because the signal for the $S>200$ mJy subsample is affected by a 'bump' at $\theta\sim0\fdg2$ 
(see Fig. \ref{fig:comp_nvss}), we obtained the amplitudes for this subsample by fitting the small- and large-scale correlation functions 
separately with power-laws $w(\theta)=B\theta^{-3.4}$ for $\theta\le0\fdg1$ and $w(\theta)=A\theta^{-0.8}$ for $\theta\ge0\fdg3$, respectively.
The measured amplitudes and their $1\sigma$ errors are listed in Table \ref{tab:wamps}. 
The values of both $B$ and $A$ are found to increase with increasing flux density limit of the subsamples. 
The increase in $B$ can be explained by the $1/\sigma$-dependency of the small-scale correlation function that is dominated by double or 
multiple component radio sources (see \se\ref{sec:first}). 
From this point onward, we will be only concerned with the amplitude $A$ that is believed to be dominated by the true cosmological clustering.  
In Fig. \ref{fig:wamps} we have plotted the 
amplitude of the cosmological \wt as a function of flux density limit. 
For comparison, we have indicated the results from the 325 MHz WENSS and 4850 MHz GB6 surveys 
\citep{reng98,reng99} by extrapolating to 1.4 GHz using a power law spectrum, $S_\nu\propto\nu^{-\alpha}$, with spectral 
index $\alpha=0.8$. Between 3 and 40 mJy the amplitude is 
approximately constant within the errors and has an (unweighted) average of $\sim1.2\times10^{-3}$. From $50-100$ mJy 
the amplitude is $\sim2\times$ higher, and it has increased by another factor of $\sim2-3$ at 200 mJy. 
These measurements indicate a trend of increasing clustering amplitude with increasing flux density limit. However, one has to keep 
in mind that the sources in the brighter subsamples are also included in the subsamples with lower limiting flux densities. 
Therefore, we also compute \wt for sources that lie in the flux {\it interval} $10<S<40$ mJy. The results are shown in Fig. \ref{fig:comp_nvss} 
together with \wt found for $S>200$ mJy. The amplitude $A=(6.6\pm1.8)\times10^{-3}$ that we measure for $S>200$ mJy 
is significantly higher than the amplitudes measured at lower flux densities. This is 
consistent with \citet{reng98} and \citet{reng99} who found $A=(11.5\pm3.5)\times10^{-3}$ 
for $S_{1.4}\ge160$ mJy in the GB6 survey and \citet{loan97} who estimated that $A$ has a value between 0.005 
and 0.015 for $S_{1.4}>100-270$ mJy from the combined 87GB and PMN surveys (Fig. \ref{fig:wamps}).  
\begin{table}[t]
\caption[]{\label{tab:wamps}Amplitudes and $1\sigma$ errors of the double power-law correlation function 
$w(\theta)=B\theta^{-3.4}+A\theta^{0.8}$ as a function of flux density limit.}
\begin{center}
\begin{tabular}[t]{lllll}
\hline
\hline
$S_{lim}$ & \multicolumn{2}{c}{NVSS} & \multicolumn{2}{c}{FIRST}\\
\hline
        & $10^6\times B$ & $10^3\times A$ & $10^6\times B$ & $10^3\times A$\\
\hline
3 mJy   & $0.7\pm0.1$ & $0.8\pm0.1$   &  $0.7\pm0.1$  &  $1.2\pm0.1$\\ 
5 mJy   & $1.0\pm0.1$ & $1.0\pm0.1$   &  $1.0\pm0.1$  &  $1.4\pm0.2$\\
7 mJy   & $1.2\pm0.1$ & $0.9\pm0.2$   &  $1.2\pm0.1$  &  $1.3\pm0.3$\\
10 mJy  & $1.5\pm0.1$ & $1.0\pm0.2$   &  $1.4\pm0.1$  &  $1.9\pm0.3$\\
20 mJy  & $2.6\pm0.2$ & $0.9\pm0.2$   & \multicolumn{2}{c}{--}\\
30 mJy  & $4.1\pm0.1$ & $1.0\pm0.2$   & \multicolumn{2}{c}{--}\\
40 mJy  & $4.4\pm0.1$ & $1.3\pm0.5$   & \multicolumn{2}{c}{--}\\
50 mJy  & $8.3\pm0.2$ & $1.9\pm0.8$   & \multicolumn{2}{c}{--}\\
60 mJy  & $6.8\pm3.0$ & $1.7\pm0.4$   & \multicolumn{2}{c}{--}\\
80 mJy  & $8.4\pm0.2$ & $1.9\pm0.7$   & \multicolumn{2}{c}{--}\\
100 mJy & $19\pm1$ & $2.2\pm0.7$   & \multicolumn{2}{c}{--}\\
200 mJy & $30\pm1$ & $6.6\pm1.8$   & \multicolumn{2}{c}{--}\\
\hline
\end{tabular}
\end{center}
\end{table}
We would like to make the following remarks:\\
(i) \citet{reng98} and \citet{reng99} measured \wt from WENSS and GB6 by excluding the first 
$5\arcmin$ and $10\arcmin$, respectively. We have used our routines to measure \wt for their catalogues as well (not shown here). 
The amplitudes and slopes we find are consistent with their values, and we find no evidence 
for a contribution of multi-component sources at the smallest angular scales allowed by these surveys.\\ 
(ii) Below 10 mJy the amplitudes for the NVSS and FIRST data are consistent with $A\simeq1.1\times10^{-3}$. 
However, at 10 mJy the amplitude is $\sim2\times$ higher for FIRST than for the NVSS. 
This is curious since the NVSS and FIRST surveys probe radio sources 
at exactly the same frequency. \citet{blake03} give a very nice demonstration (see their Fig. 3) of the most probable cause.  
The resolution of FIRST is ten times higher than that of NVSS, and therefore the average flux density of a single NVSS source is 
only equal to the {\it sum} of all its possibly resolved components in FIRST. 
Sources that appear in NVSS with integrated fluxes just above a given flux density limit can thus be missed in FIRST. 
Therefore, we consider NVSS to be more optimal than FIRST for measuring the clustering of extra-galactic radio sources. 
Furthermore, if we compute \wt for only those NVSS sources that lie in the region covered by FIRST, we find an amplitude of 
 $A=(1.7\pm0.3)\times10^{-3}$. This is consistent with the results found for the 10 mJy FIRST sample, suggesting that  
cosmic variance of clustering may be an additional factor contributing to the difference in amplitudes measured for the total NVSS area and FIRST. 
Future work might show that the region covered by FIRST is especially rich in large-scale structures.\\   
(iii) In the 200 mJy subsample we find an unexpected increase in the correlation signal at $\theta\approx0\fdg2$ (indicated by the 
connected points in Fig. \ref{fig:comp_nvss}). We investigate two possibilities. 
{\it (1) Sidelobes}: \citet{cress96} found a bump in $w(0\fdg1)$ for $S>3$ mJy sources in FIRST, and found that it was caused by sidelobe 
contamination. However, if sidelobes are responsible for boosting the correlation function at $\theta\sim0\fdg2$ in the $S>200$ mJy NVSS sample, 
these sidelobes themselves also must have minimum peak fluxes of 200 mJy. It is highly unlikely that such bright sidelobes have found their way into 
the NVSS catalogue, without being masked in \se\ref{sec:nvsssel}. Also, we have visually inspected the contour maps of 
several tens of source pairs ($S>500$ mJy) that contribute to \wt at $\theta\sim0\fdg2$. In all cases the pairs consisted of unresolved peaks without 
signs of diffuse, extended emission or side-lobe contamination. 
{\it (2) Radio galaxies with large angular sizes}: the position of the bump near the break in \wt suggests that it may somehow be 
related to the size distribution. Conveniently, \citet{lara01} have constructed a sample of 84 
large angular size ($\theta\ge4\arcmin$) radio galaxies from the NVSS at $\delta\ge+60\degr$ and a total integrated flux density of $\ge100$ mJy. 
Candidates were pre-selected by visual inspection of the NVSS maps, and confirmed or rejected following observations at higher resolution.  
If the bump is caused by $\sim12\arcmin$-sized radio galaxies, then given the 2-Mpc linear size cutoff of large radio galaxies 
\citep[see][]{schoen01}, these galaxies must lie at $z\lesssim0.1$. 
It is unlikely that such a large, relatively nearby source with, among other emission, two radio components each with a peak flux of $\ge200$ mJy 
would have been missed by their selection criteria. \citet{lara01} determined angular sizes by either measuring the maximum distance 
between $3\sigma$ contours, or by the distance between peaks at the source extremes. Also, 
sizes were measured along the 'spine' of a source if significant curvature was present. 
To investigate how many of these sources could actually contribute to \wt at $\sim12\arcmin$ we redetermine the 
angular sizes of the sources of \citet{lara01}. We find that none of these sources consists of $\ge2$ components of $\ge200$ mJy of 
$\sim12\arcmin$ separation. On the other hand, if we extrapolate the clustering power-law derived at larger scales to $\theta=0\fdg2$ we find 
that the bump translates into $\sim10\times$ the number of pairs expected. Even allowing for the much larger area of NVSS, 
the possibility that the bump is caused by large radio galaxies as in the sample of \citet{lara01} is therefore unlikely. 
     
Unfortunately, the exact origin of this feature remains unclear.  
We realize, however, that this bump is situated at a crucial angular scale for our measurements. Therefore, we have obtained the amplitudes $B$ and $A$ by 
fitting \wt on both sides of the bump with a single power-law. Under the condition that the effect that causes the bump is not responsible for enhancing \wt at 
$\theta\gtrsim0\fdg3$, this will enable us to derive an estimate for the amplitude for the cosmological clustering. At $\theta\gtrsim0\fdg3$ \wt is 
consistent with the classical $\gamma=1.8$ power-law clustering model.   
    
\subsection{Modelling the angular size distribution of radio galaxies}
\label{sec:model}

\subsubsection{The model}

The steepening of the slope of \wt at small angular scales is presumably  
related to multi-component sources spuriously enhancing  
the true clustering pair counts at small $\theta$.  
To demonstrate the reality of this assumption, we create a simple model 
for the angular size distribution of radio galaxies in the NVSS, that is able to account for this  
extra signal contributing to $w(\theta)$. We model the physical size 
distribution of sources in our $S>10$ mJy NVSS sample, and use their redshift distribution to obtain 
the angular size distribution. Because we know the angular resolution of the NVSS, this model can then be used to 
estimate the fraction of sources likely to be resolved. It is essential to separate sources that are resolved 
into a single, elongated object from sources that are resolved into a number of components, since only the latter would 
produce extra pair counts. Here, we assume that the majority of surplus pair counts arise from resolving 
the two edge-brightened radio lobes of FRII-type radio galaxies \citep[see][]{fr74}, and we estimate  
that the fraction of FRIIs at 10 mJy is $\sim40$\% from \citet{wj97} (assuming a spectral index of $\alpha=0.8$ to extrapolate to 1.4 GHz). 

Several groups have investigated the median physical sizes of FRII radio galaxies as a function of  
redshift and radio luminosity by parameterizing the linear size as $D\propto(1+z)^{-n}P^m$,
where $P$ is the radio luminosity \citep[for a review see][]{blundell99}. Results using different
samples of radio galaxies vary from no size evolution at all 
\citep[e.g.][]{nilsson93}, to size evolution depending only on redshift \citep[e.g.][]{kapahi87}, and 
size evolution depending on both redshift and luminosity with contradictory results 
\citep[e.g.][]{oort87,barthelmiley88,singal93}. We use the results of \citet{neeser95} who found the following 
linear size--redshift relation from a spectroscopically complete sample of FRII radio galaxies:  
\begin{equation}
\label{eq:evo}
D\propto(1+z)^{-1.7\pm0.5}\quad\textrm{(for $\Omega_M=1$ and $\Omega_\Lambda=0$)}, 
\end{equation} 
and remark that no intrinsic correlation was found between $D$ and $P$ ($P^m\simeq1$ with $m=0.06\pm0.09$). 
This observed linear-size evolution may be related to evolution of the confining intergalactic medium, or to evolution of 
the radio galaxy itself, but the exact underlying 
physical mechanism is unknown \citep[see][]{neeser95}. 

\begin{figure}[t]
\centering
\includegraphics[width=\columnwidth]{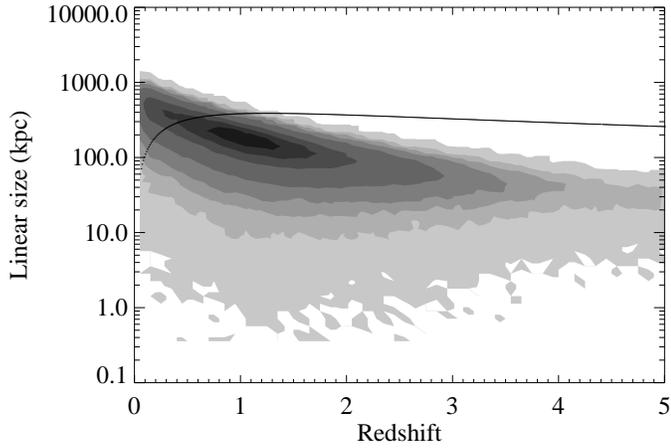}
\caption{\label{fig:sizecontours}The modeled physical size distribution of $S>10$ mJy FRII radio galaxies 
in the NVSS catalogue. The source density in the linear size-redshift plane is indicated by contours to illustrate the 
underlying redshift distribution (darker greyscales indicate higher densities). 
Sources lying above the line can, in principle, be resolved given the angular resolution of the NVSS of 
45\arcsec (FWHM).}
\end{figure}

For the purpose of our model, we place simulated sources in small redshift intervals ($\Delta z=0.01$) 
in the range $0\le z\le5$, and assume that their mean physical size evolves with redshift according to 
equation \ref{eq:evo}. We set the total number of input sources equal to the estimated number of
$S>10$ mJy FRIIs in our NVSS sample ($\sim40\% of 434,000$), and calculate the number of 
sources in each redshift interval from the redshift distribution, $N(z)$, using the formalism of \citet{dp90} (see \se\ref{sec:nz} for details). 
We then assume that in each redshift interval sizes are 
normally distributed. We take a mean size of 500 kpc and a standard deviation of 250 kpc at $z\simeq0$, chosen so 
that the resulting physical size distribution roughly resembles the distribution of projected linear sizes versus 
redshift as it is given by \citet{blundell99} for three complete samples of FRII radio galaxies from the 
3C, 6C, and 7C radio surveys. The resulting physical size distribution is shown in Fig. \ref{fig:sizecontours}, 
where we plot filled contours of 
the source density in the linear size-redshift plane to illustrate the underlying redshift distribution. 
We have also indicated the minimum physical size that is theoretically required for a source 
to become resolved as a function of redshift, given by the NVSS resolution of 45\arcsec~ (FWHM). 
We would like to remark at this point that the distribution of sizes in our model beyond redshifts of $z\sim3$ 
should not be taken too seriously as it is based on a straight extrapolation from measurements made at redshifts 
$0\lesssim z\lesssim2$, and does not take into account the fact that at these high redshifts most sources will 
be extremely young and are thus likely to be very small. However, as can be seen from Fig. \ref{fig:sizecontours}, 
our modeled size distribution falls below the NVSS resolution already at $z\sim1$. Taking smaller sizes at higher redshifts 
will have no effect on the modeled size distribution of resolved sources that we want to derive here.  

\begin{figure}[t]
\centering
\includegraphics[width=\columnwidth]{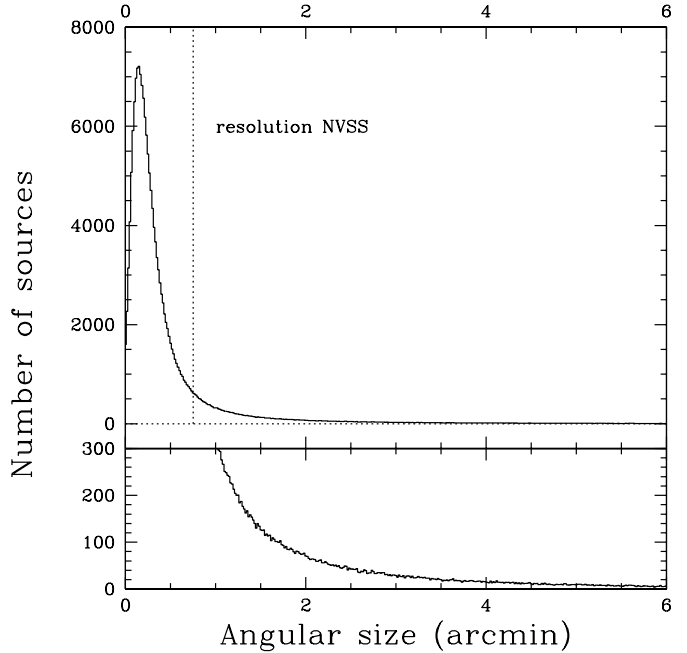}
\caption{\label{fig:angsizes}The angular size distribution for FRII radio
galaxies in the NVSS calculated from the modeled physical size distribution 
(assuming $\Omega_M=1$). The number of input
sources was chosen to match the predicted number of FRIIs in the $S>10$ mJy subsample. 
The binsize is 1\arcsec.}
\end{figure}

Assuming $\Omega_M=1$ we calculate the 
angular size distribution associated with our model. We construct 10 such models, and average them 
to get our final model of the angular size distribution of the sample. 
This model is presented in Fig. \ref{fig:angsizes}. Although the mean angular size is $\sim$10$\arcsec$ in agreement with 
\citet{condon98}, sizes are found to extend up to several arcminutes beyond the resolution 
of the NVSS (indicated by the dotted line). 

\subsubsection{Results}

\begin{figure}[t]
\centering
\includegraphics[width=\columnwidth]{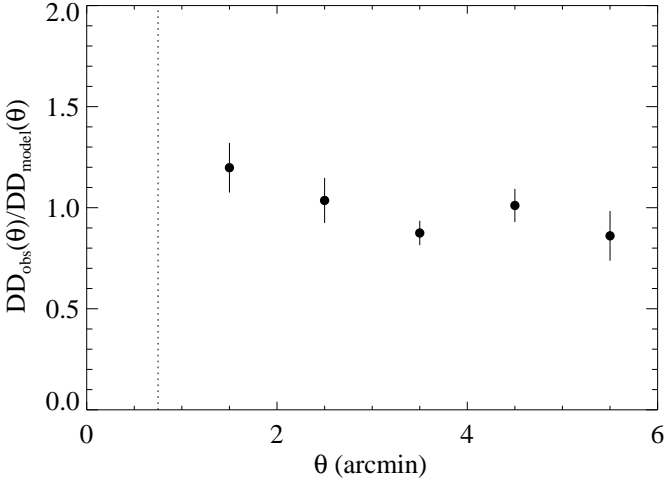}
\caption{\label{fig:modeldouble}The ratio of observed doubles to modeled doubles per distance interval. 
The angular resolution of the NVSS is indicated by the dotted line.}
\end{figure}

We now compare the number of surplus pairs expected from resolved FRII sources in the model, $DD_{mod}(\theta)$, 
to the actually measured pair counts at angular scales of $\theta\lesssim6\arcmin$. 
At these scales, the measured pair counts consist of both pair counts due to clustering and pair counts due to doubles, so 
\begin{equation} 
DD_{tot}(\theta)=DD_{gal}(\theta)+DD_{dbl}(\theta). 
\end{equation}
To extract $DD_{dbl}(\theta)$ from the total counts, $DD_{tot}(\theta)$, we calculate $DD_{gal}(\theta)$ by 
assuming that the galaxy angular correlation function as measured above the break in \wt can be extrapolated to  
 angular scales of $\theta\lesssim6\arcmin$: 
\begin{equation}
w_{gal}(\theta)=1.0\times10^{-3}\theta^{-0.8}= DD_{gal}(\theta)\cdot F(\theta)-1,
\end{equation}
where $F(\theta)=4\cdot RR(\theta)/[DR(\theta)]^2$, the part of the Hamilton estimator that is
relatively independent of the presence of doubles.
Since we now know both $DD_{tot}(\theta)$ and $DD_{gal}(\theta)$, we can
subtract them to get a measure of the counts arising from the double
sources: $DD_{dbl}(\theta)$. The final step is to rebin the modeled number of pair 
separations $DD_{mod}(\theta)$ in order to match the binning scheme of
$DD_{dbl}(\theta)$. Fig. \ref{fig:modeldouble} shows the
ratio of the observed doubles to the modeled doubles per distance interval. 
The errors in the observed counts are estimated from the 
$1\sigma$-error in the amplitude of $w(\theta)$. The errors in the modeled pair counts are estimated 
by allowing a 10\% error in the estimated fraction of FRIIs in the NVSS.  
We conclude that: {\it a model in which the small-scale angular correlation function steepens due to 
resolving FRII radio galaxies into two distinct knots of radio emission is in good agreement with the measurements presented in 
Fig. \ref{fig:10log}}.   

Several remarks that can be made are the following:\\
(i) The size distribution of radio sources at the mJy level is still largely unconstrained.  
Recently, however, \citet{lara01} presented a new sample of large radio galaxies (LRGs) selected from the NVSS. 
In the region $\delta\ge+60\degr$ they found $\sim$80 radio galaxies with apparent angular sizes larger than $4\arcmin$ and 
total flux density greater than 100 mJy. If we roughly extrapolate our model to their sensitivity and correct for the area we successfully 
predict the number of FRIIs in the range $4\arcmin\lesssim\theta\lesssim6\arcmin$. However, in this interval one third of the sample of \citet{lara01} 
consists of FRIs, while the model only uses FRIIs to estimate the number of surplus pairs expected. The model could be refined by decreasing the 
fraction of resolved FRIIs to also allow a contribution from large FRIs.\\  
(ii) The model allows objects to be either single or double sources, 
although visual inspection of NVSS contour maps shows that sources are sometimes split into three or even more 
components. Therefore, we may expect an extra amount of spurious pair counts on top of the counts due to classical 
double radio sources. This may become increasingly important with increasing flux density limit.\\
(iii) The model predicts a fraction of resolved sources in NVSS of $\sim$0.07, in rough agreement with 
the value of $\sim$0.05 predicted by \citet{condon98}.

The simple model allows us to explore the general  
relations between the physical size distribution of radio galaxies
and \wt at small angular scales. Although our crude method is successful in 
reproducing the observations, it relies on a number of assumptions that are not easily verified 
from the data currently in literature. Radio sources come in a wide variety of sizes ranging from 
$<1$ kpc for the class of gigahertz peaked spectrum sources (GPS), to 1--20 kpc for the compact steep 
spectrum sources (CSS), $>20$ kpc for FRI- and FRII-type radio galaxies, and $>1$ Mpc for giant radio galaxies  
\citep{fanti90,odea91,blundell99,schoen01}. Evidently, the distribution of linear sizes of radio
sources are very complex, and will remain an important subject for future studies.  
As we have shown, the angular correlation function can be used to put constraints on the size distribution of 
large radio galaxies. However, perhaps more ideal would be to make a statistical redshift sample of {\it all} radio source 
pairs within some angular distance interval, and then take high resolution radio observations to constrain the 
numbers of intrinsic doubles in that sample.  

\section{The spatial clustering of NVSS souces}
\label{sec:nz}

\subsection{The redshift distribution}

At the mJy level and higher it is standard practice to compute redshift distributions 
using the \citet{dp90} radio luminosity functions (RLFs). These authors have constructed a range of  
model luminosity functions using spectroscopically complete samples from several radio surveys 
at different frequencies. Using a free-form modelling approach they found a number of smooth functions that were consistent with 
the data. In addition, they attempted two models of a more physical nature by assuming pure luminosity evolution (PLE) and 
luminosity/density evolution (LDE) to describe the RLF. The total ensemble is expected to agree well at those 
luminosities and frequencies at which they are best constrained by the data, while uncertainties in the extrapolation 
of each of these models to those regions that are less constrained by the data may be reduced by taking the ensemble as 
a whole. We compute redshift distributions, $N(z)$, for each flux-limited subsample using the free-form models $1-4$ and 
the PLE/LDE models for the combined population of flat ($\alpha=0$, $S_{\nu}\propto\nu^{\alpha}$) and steep ($\alpha=-0.8$) 
spectrum radio sources given by \citet[][taking the MEAN-$z$ data from their appendix C]{dp90} from
\begin{eqnarray}
\frac{dN(z)}{dz}&=&\frac{dV(z)}{dz}\times\int_{P_{low}(z)}^\infty \Phi_{i}(P,z)dP,\\
P_{low}(z)& =& x(z)^2 \left(\frac{S}{(1+z)^{1-\alpha}}\right)\left(\frac{2.7~ GHz}{\nu}\right)^\alpha,\nonumber
\end{eqnarray}
where $V(z)$ is the comoving volume, $\Phi_{i}(P,z)$ is the model RLF, $x(z)$ the comoving distance, 
$S$ the limiting flux density of the subsample, and $\nu$ the frequency of FIRST/NVSS. We note that $N(z)$ 
is independent of cosmology as long as the calculations are carried out in the cosmology used to construct the RLFs 
(i.e. $\Omega_M=1.0$ and $H_0=50$ km s$^{-1}$ Mpc$^{-1}$).  

\begin{figure}[t]
\includegraphics[width=\columnwidth]{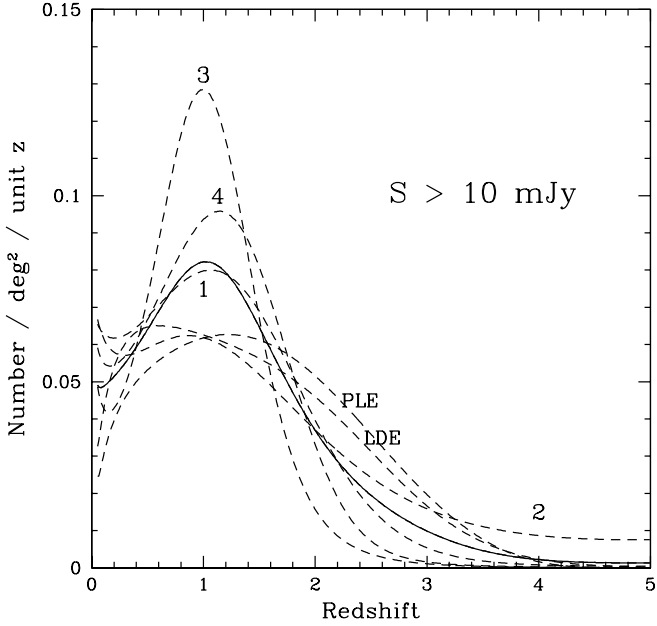}
\caption[]{\label{fig:dp90a}Dashed lines show the redshift distributions for $S_{1.4}>10$ mJy, 
computed from the free-form models $1-4$, the pure luminosity evolution model $(PLE)$ and 
the luminosity/density evolution model $(LDE)$ of \citet{dp90} 
(see text for details). The average of the six different models is indicated by the solid curve.}
\end{figure}

\begin{figure}[t]
\includegraphics[width=\columnwidth]{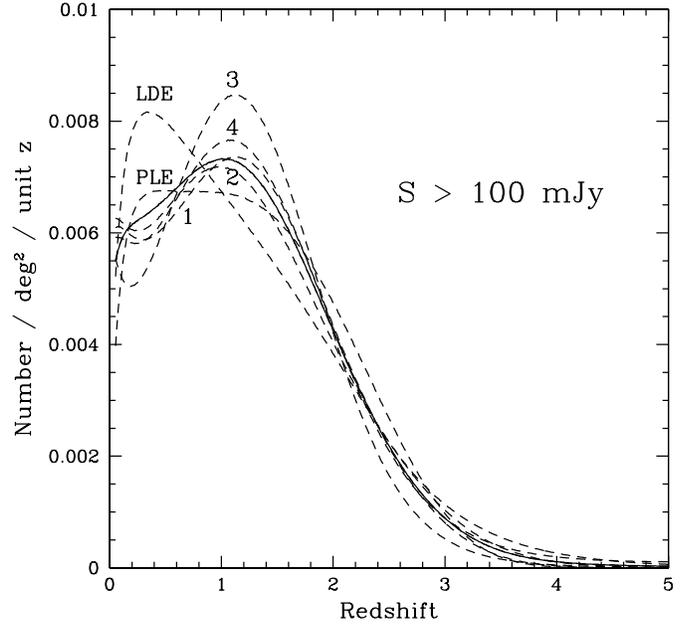}
\caption[]{\label{fig:dp90b}The redshift distributions for $S_{1.4}>100$ mJy. See the caption of Fig. \ref{fig:dp90a}  
for details.}
\end{figure}

Figs. \ref{fig:dp90a} and \ref{fig:dp90b} show the redshift distributions for $S>10$ mJy  
and $S>100$ mJy, respectively. We calculate the average of the six different models (indicated by the solid curve), 
which will be our best estimate of $N(z)$ use in the analysis below (the same method was used for the $N(z)$ applied to the 
model of the angular size distribution described in \se\ref{sec:model}). It is important to keep in mind that the functional 
form of $N(z)$ remains virtually unchanged from $3-200$ mJy. Over this range in flux densities the RLFs represent a    
broad redshift distribution with a peak around $z\sim1$, indicating the very large median redshift that is 
generally probed by radio surveys. 

\subsection{The spatial correlation function}
\label{sec:spatial}

Given the amplitudes of \wt determined in \se\ref{sec:results} we can use the cosmological Limber equation to estimate  
the spatial correlation length, $r_0$, by deprojecting \wt into the spatial 
correlation function, $\xi(r)$ using the redshift distribution and cosmology (e.g. Peebles 1980, \S56)\nocite{peebles80}. 
We consider two cosmological models: 
a flat, vacuum dominated, low-density Universe ($\Lambda$CDM; $\Omega_M=0.3$, $\Omega_\Lambda=0.7$), 
and an Einstein-de Sitter model Universe ($\tau$CDM; $\Omega_M=1.0$, $\Omega_\Lambda=0$). 
We use $H_0=100h$ km s$^{-1}$ Mpc$^{-1}$.  

We assume an epoch dependent power-law spatial correlation function of the form
\begin{equation}
\xi(r_p,z)=\left(\frac{r_p}{r_0}\right)^{-\gamma}(1+z)^{-(3+\epsilon)},
\end{equation}
where $r_p$ is the proper distance, $r_0$ is the spatial correlation length\footnote{Note that the spatial 
correlation length is {\it not} a physical lengthscale in the space distribution of galaxies. 
It is just defined as that length at which $\xi(r)$ is unity 
(i.e. the chance of finding a galaxy at the distance $r_0$ from another galaxy is 
twice the Poissonian chance).} at $z=0$, and $\epsilon$ parameterizes the redshift evolution 
of the clustering. To express $\xi(r_p,z)$ in terms of comoving coordinates $r_c=r_p\times(1+z)$, we write:
\begin{equation}
\xi(r_c,z)=\left(\frac{r_c}{r_0}\right)^{-\gamma}(1+z)^{\gamma-(3+\epsilon)},
\end{equation} 
which can be written as 
\begin{equation}
\label{eq:r0}
\xi(r_c,z)=\left(\frac{r_c}{r_0(z)}\right)^{-\gamma},~ r_0(z)=r_0(1+z)^{1-\frac{3+\epsilon}{\gamma}},
\end{equation} 
where $r_0(z)$ is the (comoving) correlation length measured at $z$.
In a flat model Universe, the cosmological Limber equation can be expressed as follows \citep[see e.g.][]{peebles80}:
\begin{eqnarray}
w(\theta) & = & A\theta^{1-\gamma} = \sqrt{\Omega_M}\left(\frac{r_0H_0}{c}\right)^\gamma\theta^{1-\gamma}H_\gamma\times\\
          &   & \frac{\int_0^\infty dz~N(z)^2(1+z)^{\gamma-3-\epsilon}x^{1-\gamma}Q(z)}{\left[\int_0^\infty dz~N(z) \right]^2},\nonumber
\end{eqnarray}
with
\begin{eqnarray}
Q(z) & = & \left[(1+z)^3 + \Omega_M^{-1}-1 \right]^{0.5},\\
x(z) & = & \frac{1}{\sqrt{\Omega_M}}\int_0^z \frac{dz}{Q(z)},\nonumber\\
H_\gamma &=&\Gamma\left(\frac{1}{2}\right)\Gamma\left(\frac{\gamma-1}{2}\right)\Gamma\left(\frac{\gamma}{2}\right)^{-1},\nonumber
\end{eqnarray}
and using the approximation that angles are small ($\theta\ll1$). We calculate $N(z)$ for each subsample.  

The evolution parameter $\epsilon$ can represent a variety of clustering models. Three important cases are the 
following \citep[see][]{phil78,kundic97}. {\it (1) The stable clustering model} ($\epsilon=0$): 
if galaxy clustering is gravitationally bound at small scales, then clusters have fixed physical sizes 
(i.e. they will neither contract nor expand) and will have a correlation function that decreases with redshift 
as $(1+z)^{-1.2}$. {\it (2) The comoving clustering model} ($\epsilon=\gamma-3$): galaxies and clusters expand with the Universe, 
so their correlation function remains 
unchanged in comoving coordinates. This case applies well to a low density Universe 
where there is not enough gravitational pull to counterbalance expansion,  
and implies that structures have formed very early. 
{\it (3) The linear growth model} ($\epsilon=\gamma-1$): clustering grows as expected under linear perturbation 
theory.

\begin{figure}[t]
\centering
\includegraphics[width=\columnwidth]{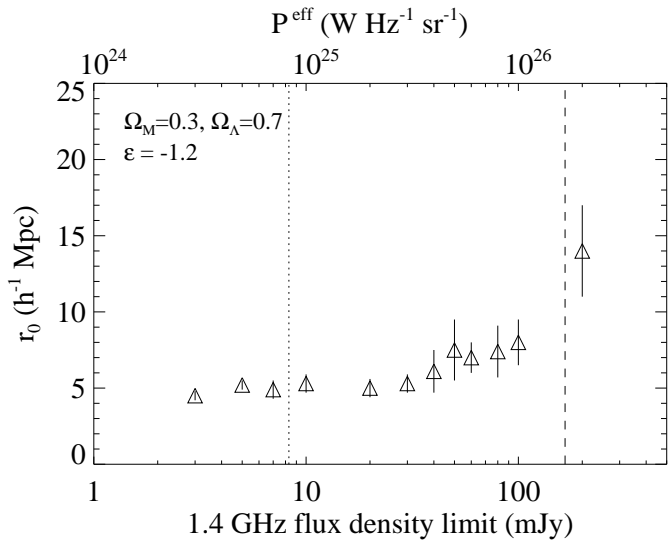}
\caption{\label{fig:r0amps}Spatial correlation lengths and $1\sigma$ errors derived from the cosmological 
$w(\theta)$ of the NVSS, assuming an evolution parameter 
$\epsilon=-1.2$, and the $\Lambda$CDM model Universe. The dotted line indicates the flux density limit at 
which FRI- and FRII-type radio sources contribute roughly equally to 1.4 GHz radio source counts. The dashed line indicates the flux density limit above which the contribution of FRIIs is $\gtrsim75\%$.
The top axis indicates the effective radio luminosity as a function of flux density limit.}
\end{figure}

Studies of the spatial clustering properties of radio-quiet quasars indicate that the clustering history 
of active galaxies, unlike that of normal galaxies, is best characterized using a {\it negative} value for $\epsilon$.
\citet{kundic97} measured the high-redshift quasar-quasar correlation function from the 
Palomar Transit Grism Survey, and found no evidence for a decrease in the correlation amplitude of quasars 
with redshift. Moreover, he found that $\xi_{qq}(z>2)/\xi_{qq}(z<2)\simeq1.8$, suggesting an even higher 
amplitude at higher redshifts. Similarly, \citet{croom01} find almost no evolution in clustering strength 
for quasars taken from the 2dF QSO Redshift Survey out to $z\simeq2.5$.  
\begin{table*}[t]
\caption[]{\label{tab:r0amps}Present-day spatial correlation lengths and $1\sigma$ errors derived from the 
galaxy \acf ($\gamma=1.8$) of the NVSS as a function of flux density limit. Listed are the results found using two 
different cosmological models and two different values for the evolution parameter $\epsilon$ (see text for details).}
\begin{center}
\begin{tabular}[t]{ccccc}
\hline
\hline
 &\multicolumn{2}{c}{$\epsilon=-1.2$} & \multicolumn{2}{c}{$\epsilon=0$}\\
\hline
 $S_{low}$ & $\tau$CDM & $\Lambda$CDM & $\tau$CDM & $\Lambda$CDM\\
 & $r_0$ ($h^{-1}$ Mpc) & $r_0$ ($h^{-1}$ Mpc) & $r_0$ ($h^{-1}$ Mpc) & $r_0$ ($h^{-1}$ Mpc)\\
\hline
3 mJy   &  $3.2\pm0.2$ & $4.5\pm0.3$ & $4.8\pm0.3$  & $6.3\pm0.5$\\ 
5 mJy   &  $3.7\pm0.2$ & $5.2\pm0.3$ & $5.6\pm0.3$  & $7.5\pm0.4$\\
7 mJy   &  $3.5\pm0.5$ & $4.9\pm0.6$ & $5.3\pm0.7$  & $7.2\pm0.9$\\
10 mJy  &  $3.7\pm0.4$ & $5.3\pm0.6$ & $5.7\pm0.7$  & $7.8\pm0.9$\\
20 mJy  &  $3.5\pm0.4$ & $5.0\pm0.6$ & $5.5\pm0.7$  & $7.5\pm1.0$\\
30 mJy  &  $3.7\pm0.4$ & $5.3\pm0.6$ & $5.8\pm0.7$  & $7.9\pm0.9$\\
40 mJy  &  $4.3\pm1.0$ & $6.1\pm1.4$ & $6.7\pm1.6$  & $9.1\pm2.2$\\
50 mJy  &  $5.3\pm1.4$ & $7.5\pm2.0$ & $8.2\pm2.2$  & $11.2\pm2.9$\\ 
60 mJy  &  $5.0\pm0.7$ & $7.0\pm1.0$ & $7.7\pm1.1$ & $10.4\pm1.4$\\
80 mJy  &  $5.3\pm1.2$ & $7.4\pm1.7$ & $8.1\pm1.8$  & $10.9\pm2.5$\\
100 mJy &  $5.7\pm1.1$ & $8.0\pm1.5$ & $8.7\pm1.7$ & $11.6\pm2.2$\\
200 mJy &  $10.6\pm1.8$& $14\pm3.0$  & $15.4\pm2.6$ & $19.8\pm3.4$\\
\hline
\end{tabular}
\end{center}
\end{table*}
Therefore, we opt for evolution model 2 
(i.e. constant clustering in comoving coordinates), which implies $\epsilon=-1.2$ for $\gamma=1.8$. 
In table \ref{tab:r0amps} we list the results obtained using this model for the two 
different cosmological models. For comparison, we also indicate the results using the stable clustering 
model ($\epsilon=0$). For $\epsilon=0$ the present-day correlation length is $\sim1.4$ times higher than for 
$\epsilon=-1.2$ in both cosmologies. However, given the strong peak in the redshift distribution at $z\sim1$, we are effectively 
measuring clustering at $z\sim1$. Calculating $r_0(z\sim1)$ in the case of stable clustering using Eq. \ref{eq:r0} yields 
a value that is only $\sim1.1$ times lower than $r_0(z\sim1)=r_0$ in the case of $\epsilon=-1.2$. 
Therefore, the value of $r_0(z\sim1)$ is relatively independent of the exact value of $\epsilon$. 
The results for the $\epsilon=-1.2$ ($\Lambda$CDM) case are presented in Fig. \ref{fig:r0amps}. 
We find an approximately constant spatial correlation length of $\simeq6.0$ $h^{-1}$ Mpc from 3--40 mJy, 
compared to $\simeq14$ $h^{-1}$ Mpc at 200 mJy. 

As we have shown, the possibility that the observed flux-dependency of the clustering is just an effect of projection can be ruled out, 
since the shape of the redshift distribution is relatively constant with flux over several orders of magnitude 
(at least above $\sim1$ mJy). This automatically implies that the average radio power of the subsamples increases with flux density 
(indicated by the top axis of Fig. \ref{fig:r0amps}). An alternative explanation was therefore suggested by \citet{reng98} and \citet{reng99} 
based on their measurements of the clustering of radio sources in the WENSS and GB6 surveys. They concluded that the clustering signal could change 
as a function of flux density if relatively low and high power radio galaxies represent different spatial structures 
at a similar epoch ($z\sim1$). Taking the predicted population mix of radio sources from \citet{wj97}, 
we find that for $S_{1.4}>10$ mJy the fractions of FRIs and FRIIs are about equal. However, for $S_{1.4}>100$ mJy 
the fraction of FRIIs is more than $\sim75$\%. Given the fractional changes of the source populations with flux density limit, 
the clustering amplitudes measured are very well matched by a scenario in which 
{\it the clustering of powerful radio sources (mostly FRII) and average power radio sources (FRI/FRII) are intrinsically different, with 
FRIIs being more strongly clustered at $z\sim1$ than the radio galaxy population on average.}

As pointed out by \citet{reng98} and \citet{reng99} the large difference in observing frequencies and sensitivities 
of WENSS and GB6 (the limiting 1.4 GHz flux densities probed by these surveys correspond 
to 10 mJy for WENSS and 70 mJy for Greenbank, respectively) only allowed them to make a comparison between the results, whereas 
the detection of the inferred flux-dependency of $r_0$ within a {\it single} survey would be highly desirable. 
Our analysis of the clustering in the single large-area, intermediate-frequency NVSS survey is in agreement with their conclusions. 

\section{Discussion}

\subsection{Clustering measurements from literature}

We start this section by making a survey of other clustering measurements from literature. However, readers may 
wish to skip directly to \se\ref{sec:evolution} for a discussion on these measurements and the results presented 
in this paper in their cosmological context.  

In order to compare results from different studies, all values taken from literature were 
converted assuming a fixed slope $\gamma=1.8$ by setting $r_{0,1.8}=({r}_{0,\gamma})^{\gamma/1.8}$. 
All correlation lengths are expressed in comoving units, and we have transformed 
all values to a $\Lambda$CDM cosmology \citep[see][]{mag00}. 
Please note that the list given below is not complete, and the reader 
is kindly invited to consult the individual papers and the references therein for further information.

\subsubsection{Clusters}  

Estimates of the correlation length of rich Abell clusters are given 
by \citet{bahcall83} and \citet{postman92} who found $r_0=24\pm9$ $h^{-1}$ Mpc. 
\citet{lahav89} found $r_0=21\pm7$ $h^{-1}$ Mpc from an all-sky sample of the brightest X-ray clusters, and  
\citet{dalton94} and \citet{croft97} found $r_0=19\pm5$ $h^{-1}$ Mpc and $r_0=16\pm4$ $h^{-1}$ Mpc, respectively, for 
clusters selected from the APM Galaxy Survey. Recently, \citet{gonzo02} measured 
the correlation length of distant clusters in the Las Campanas Distant Cluster Survey and found a 
correlation length of $24.8\pm4.5$ $h^{-1}$ Mpc at $\bar{z}=0.42$. 

Different studies may have sampled clusters of different degrees of richness, which can account for most of the scatter 
in the reported values. In general, however, all results are consistent with clusters being the most strongly 
clustered objects known in the Universe.

\subsubsection{Optically-selected ordinary galaxies and IRAS galaxies}

Bright early-type galaxies are found to have a strongly clustered distribution 
in the local Universe. \citet{willmer98} find $r_0=6.8\pm0.4$ $h^{-1}$ Mpc for local $L\gtrsim L_*$ 
ellipticals, and \citet{guzzo97} measure a considerably higher $r_0=11.4\pm1.3$ $h^{-1}$ Mpc for a sample 
of similar galaxies. Although these results are only consistent with each other at the $3\sigma$ level, 
the latter sample contains a higher fraction of local clusters, presumably responsible for boosting the $r_0$.
The dependence of galaxy clustering on luminosity and spectral type has been studied using 
the ongoing 2 degree Field Galaxy Redshift Survey (2dFGRS). \citet{norberg02} find $r_0=11.8\pm1.6$ $h^{-1}$ Mpc 
for the brightest early-type galaxies in the 2dFGRS. Moreover, they find a strong dependence of clustering strength on 
luminosity, with the amplitude increasing by a factor of $\sim2.5$ between $L_*$ and $4L_*$. 
The ordinary population of galaxies has been found to be less strongly clustered than 
the population consisting of local (bright) ellipticals: \citet{loveday95} find $r_0=4.7\pm0.2$ $h^{-1}$ Mpc 
from the APM survey. At higher redshifts, the clustering strength in a sample of faint $K$-selected galaxies 
with minimum rest-frame luminosities of $M_K=-23.5$, or about $0.5L_*$, is found to be fairly rapidly declining with redshift: 
\citet{carlberg97} find $r_0=3.3\pm0.1$ $h^{-1}$ Mpc, $r_0=2.3\pm0.2$ $h^{-1}$ Mpc, 
$r_0=1.6\pm0.2$ $h^{-1}$ Mpc, and $r_0=1.2\pm0.2$ $h^{-1}$ Mpc, at $\overline{z}=0.34$, $\overline{z}=0.62$, 
$\overline{z}=0.97$, and $\overline{z}=1.39$, respectively. \citet{carlberg00} present measurements on a sample 
of $L\sim L_*$ galaxies up to $z\approx0.6$ and find a much milder decline from $r_0=5.1\pm0.1$ $h^{-1}$ Mpc 
at $\overline{z}=0.10$ to $r_0=4.2\pm0.4$ $h^{-1}$ Mpc at $\overline{z}=0.59$.

Clustering of the local population of IRAS-selected galaxies is best fit by $r_0=3.4\pm0.2$ 
$h^{-1}$ Mpc \citep{fisher94}.

\subsubsection{Extremely red objects (EROs)}

Several recent studies indicate that the comoving correlation length of early-type galaxies undergoes 
little or no evolution from $0\lesssim z\lesssim1$. Evidence for this is provided by the 
clustering of extremely red objects, a population of galaxies 
having very red optical to infrared colors ($R-K_s>5$). These red colors are consistent with them being 
either old, passively evolving elliptical galaxies, or strongly dust-enshrouded starburst galaxies at $z\sim1-1.5$. 
Indeed, further observations have confirmed that both classes are present in the ERO population 
\citep[e.g.][]{dunlop96,cimatti98,dey99,liu00}. \citet{daddi01} have recently 
embarked on a study of the spatial clustering of a large sample of $L\gtrsim L_*$ EROs at $z\sim1$, 
and found a large correlation length 
of $r_0=12\pm3$ $h^{-1}$ Mpc. In \citet{cimatti02} the results are presented   
involving the EROs that were identified in a large flux limited redshift survey of $\sim500$ galaxies with $K\le20$. 
The derived fraction of early-type EROs from that sample is $50\pm20\%$, while there is an increasing contribution 
of dusty star-forming EROs at faint magnitudes. Therefore, \citet{daddi02} have attempted to analyse separately 
the spatial clustering of EROs from both categories by studying the frequency of close pairs. They find that the 
comoving correlation length of the dust-enshrouded starbursts is constrained to be less than $r_0=2.5$ $h^{-1}$ Mpc, 
while the old EROs are clustered with $5.5\lesssim r_0\lesssim16$ $h^{-1}$ Mpc. This is consistent with the 
value reported earlier in \citet{daddi01}, which is still valid as a lower limit for the clustering of 
early-type EROs based on the argument that the much less clustered dusty star-forming EROs only dilute the 
clustering signal coming from the ellipticals in this sample \citep[see also][]{roche02}. 
Furthermore, \citet{mc01} have identified a large sample 
of such faint red galaxies as being consistent with mildly evolved early-type galaxies at $z\sim1.2$. 
They find a clustering strength of $r_0=9.5\pm1$ $h^{-1}$ Mpc. 

\subsubsection{Radio galaxies}

The results on the spatial clustering of radio sources at $z\sim1$ presented in this paper indicate 
that $r_0$ depends on radio luminosity in such a way that very luminous (FRII) radio galaxies cluster more strongly 
than the total population of radio galaxies (both FRI and FRII) on average, reminiscent of a similar luminosity trend found for samples of optically-selected 
 galaxies. We roughly construct two radio luminosity bins from our measurements by comparing the $r_0$ found for 3--40 mJy to the $r_0$ found 
for the 200 mJy subsample. We find $r_0\simeq6\pm1$ $h^{-1}$ Mpc$^{-1}$ for the relatively low power bin ($P_{1.4}\sim10^{24-25}$ W Hz$^{-1}$ sr$^{-1}$), 
and $r_0\simeq14\pm3$ $h^{-1}$ Mpc$^{-1}$ for the high power bin ($P>10^{26}$ W Hz$^{-1}$ sr$^{-1}$). 

\subsubsection{Optically-selected quasars}
 
\citet{croom01} have determined the correlation length of quasars (QSOs) using 10,558 quasars  
taken from the 2dF QSO Redshift Survey. They find that QSO clustering 
appears to vary little with redshift, with $r_0=4.9\pm0.8$ $h^{-1}$ Mpc at 
$\overline{z}=0.69$, $r_0=2.9\pm0.8$ $h^{-1}$ Mpc at $\overline{z}=1.16$, $r_0=4.2\pm0.7$ $h^{-1}$ Mpc 
at $\overline{z}=1.53$, $r_0=5.3\pm0.9$ $h^{-1}$ Mpc at $\overline{z}=1.89$, and $r_0=5.8\pm1.2$ $h^{-1}$ Mpc 
at $\overline{z}=2.36$.

\subsubsection{Lyman-break galaxies}

Lyman-break galaxies (LBGs) are found to be associated with star-forming galaxies at $z\sim3$, 
with comoving correlation lengths of $r_0=3.3\pm0.3$ $h^{-1}$ Mpc \citep{adelberger00}, and $r_0=3.6\pm1.2$ $h^{-1}$ Mpc 
\citep{porciani02}. \citet{ouchi01} find $r_0=2.7\pm0.6$ $h^{-1}$ Mpc 
for a sample of LBGs at $z\sim4$.   

\subsection{Clustering evolution}
\label{sec:evolution}

\subsubsection{The clustering of massive ellipticals at $z\sim1$}

\begin{figure*}[t]
\centering
\includegraphics[width=\textwidth]{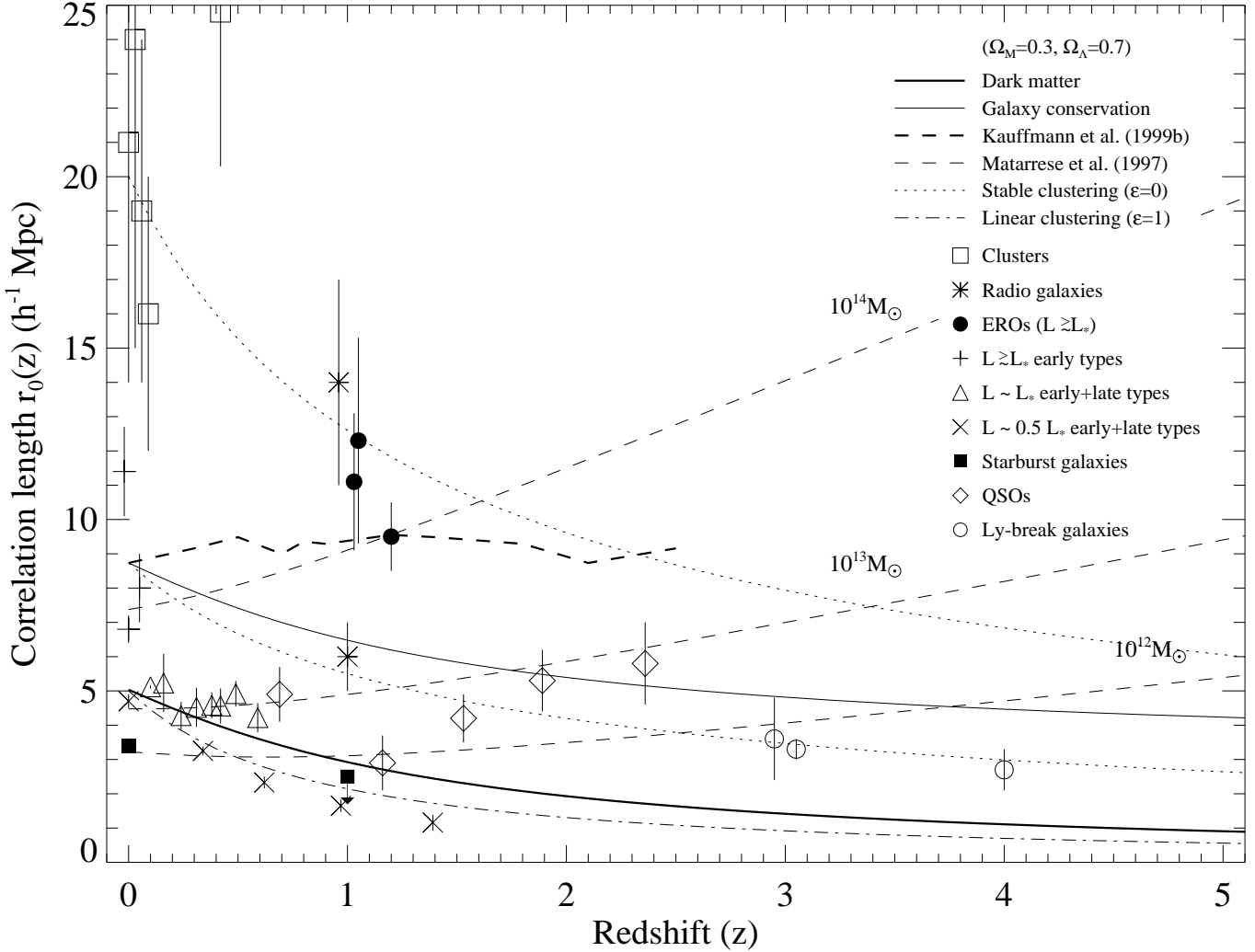}
\caption{\label{fig:zoo}
The redshift evolution of galaxy clustering in a $\Lambda$CDM Universe. See the text for references to data taken from literature.  
Lines represent the following models: (i) stable clustering ($\epsilon=0$) normalized to $r_0$ of local ellipticals and clusters  
({\it dotted lines}), (ii) linear clustering ($\epsilon=1$) normalized to 5 h$^{-1}$ Mpc ({\it dot-dashed line}), 
(iii) clustering of the dark matter ({\it thick solid line}, from \citet{jenkins98}, see also \citet{moustakas02} for a 
useful parameterization), 
(iv) galaxy conservation model normalized to $r_0$ of local ellipticals ({\it thin solid line}, see \citet{fry96}), 
(v) hierarchical model for clustering evolution of early-type galaxies normalized to $r_0$ of local ellipticals ({\it thick dashed line},
from \citet{kauffmann99b}), and (vi) clustering evolution as a function of dark matter halo masses with $\mathcal{M}_{min}=10^{12-14}M_\odot$ 
({\it thin dashed lines}, from \citet{matarrese97}). A nice representation of this figure showing actual images of the various objects rather 
than symbols can be found at our website: http://www.strw.leidenuniv.nl/\~{}overzier/r0.html.}
\end{figure*}

In Fig. \ref{fig:zoo} we present an overview of the evolution of galaxy clustering, 
as it follows from the broad variety of observational results summarized above.  
The $r_0$ that we measure for the brightest radio sources at $z\sim1$ is comparable to the $r_0$ measured   
for bright ellipticals locally, and $\sim2\times$ higher than the $r_0$ measured for relatively faint radio sources  
and quasars, suggesting that they are considerably more biased and probably probe spatial structures 
 associated with strongly clustered, massive objects. This does not come totally unexpectedly, as there is a range of 
observational evidence in support of this result. \citet{best98} found that powerful 3CR radio 
galaxies are mostly associated with massive galaxies at $z\sim1$, and at high ($z\sim1$) and very high 
($z\gtrsim2$) redshifts the most luminous (i.e. FRII-type) radio sources are found in very dense environments 
associated with forming clusters. This is based on for example the presence of large X-ray halos \citep{crawford96}, 
excesses of companion galaxies \citep{mc88,rott96,nakata01}, and excesses of Ly$\alpha$ emitters  around powerful radio 
sources \citep{kurk00,venemans02}. Furthermore, most very high redshift radio galaxies ($z>2$) are 
surrounded by giant halos of emission line gas \citep[e.g.][]{rott99,carlos00}, and some have very clumpy 
morphologies suggestive of massive forming systems \citep[e.g.][]{laura99,laura00}. Using HST/NICMOS observations, 
\citet{laura01} have found a number of radio galaxies at $z\sim2$ having morphologies that are represented well by a de 
Vaucouleurs profile, consistent with them being elliptical galaxies or proto-galaxy bulges. 

As argued by \citet{best99}, powerful radio sources must rely on (i) a plentiful supply of gas to fuel a 
supermassive blackhole that can drive the AGN activity, and (ii) a dense surrounding medium able to contain the radio lobes. 
These environments are indeed expected to be found in the gas-rich galaxy clusters at high redshift, additionally  
supporting the conclusion that high redshift FRIIs are associated with strongly clustered, massive objects. 
One may argue that this conclusion somewhat contradicts the fact that low redshift FRIIs are primarily found to be 
situated in small, isolated galaxy groups, and not in the centers of large clusters \citep{butcher78,hill91}. 
This, however, can easily be explained by considering that the local analogs of the gas-rich cluster environments 
that are suitable for producing powerful FRIIs at high redshifts, are found in relatively small galaxy groups, 
and not in the gas-depleted centers of local rich clusters \citep{reng98}. 
 
Interestingly, we find that both EROs and powerful radio galaxies are strongly clustered with $r_0\gtrsim10$ $h^{-1}$ Mpc at $z\sim1$.  
\citet{willott01a} suggested that high-redshift radio galaxies and EROs could be identical galaxies seen at different stages of 
their evolution, based on their findings of ERO-like host galaxies for a number of radio galaxies from the 7C Redshift Survey. 
This, of course, would be highly consistent with the belief that both radio galaxies and EROs may be the progenitors of local 
bright ellipticals. They conclude that the density of radio sources with minimum radio luminosities   
of log$_{10}$ $P_{151}=24$ W Hz$^{-1}$ sr$^{-1}$ is consistent with a model in 
which all EROs go through a relatively short period of AGN activity, forming a radio galaxy somewhere between $z=2$ and $z=1$. 

However, if {\it all} EROs are radio 
galaxies at some stage, their highly clustered spatial distribution should be reflected in the spatial distribution of 
the radio galaxies. Fig. \ref{fig:zoo} shows that the clustering of EROs and radio galaxies is 
consistent only for those galaxies with radio luminosities of log$_{10}$ $P_{1400}\gtrsim26$ W Hz$^{-1}$ sr$^{-1}$. 
The surface density of such radio sources in the redshift range $1<z<2$ in the NVSS is $\sim2\times10^{-4}$ arcmin$^{-2}$, 
while the surface density of EROs 
having $K_s\le19$ and $R-K_s>5$ is $\sim0.5$ arcmin$^{-2}$ \citep{daddi01}. If we take the fraction of old ellipticals 
among EROs to be $\sim70$\% \citep{cimatti02}, then only $\sim0.06\%$ of these EROs are currently observed in their radio-loud 
phase. However, because the typically assumed AGN lifetimes are short compared to the cosmological time-scale from $z=2$ and $z=1$ 
($t_{z=2-1}\simeq3.5$ Gyr for $\Omega_M=0.3$, $\Omega_\Lambda=0.7$), the number of EROs that could undergo a radio-loud phase 
is $\sim2-20\%$ (assuming $t_{AGN}\simeq10^{7-8}$ yr.). 
These fractions can be increased significantly if, for example, we select EROs that are much redder: the density of 
EROs having $R-K_s>6$ is a factor of $\sim10$ lower compared to $R-K_s>5$ \citep{daddi01}, giving $\sim14-140\%$. 
It may be clear from the above that the unification of EROs and radio galaxies, although tempting, relies on a number of issues 
that have not yet been resolved. Further study of the luminosities, colors and morphologies of radio galaxy hosts, as well as the 
cluster environments of EROs may be expected to provide important clues for constraining this scenario.        

\subsubsection{Comparison with theoretical predictions}

Linear ($\epsilon\sim1$, dot-dashed line) or stable ($\epsilon=0$, dotted line) clustering 
evolution models have been found to best fit the measurements of ordinary, optically-selected 
galaxies at $z\lesssim1$ \citep[e.g.][and references therein]{carlberg97,carlberg00,mccracken01}. 
However, as Fig. \ref{fig:zoo} shows, these models do not provide a good description for the evolution of 
massive early type galaxies as inferred from the meassurements of local bright ellipticals 
and FRII radio galaxies and EROs at $z\sim1$. Adjusting these models  
to the measurements would either require $z\sim1$ massive ellipticals to have a correlation length around $6-7$ $h^{-1}$ Mpc, 
or local bright ellipticals to have a correlation length of the order of that of local clusters, far greater than observed. 
For these galaxies, the current measurements require a model that predicts relatively constant clustering in comoving coordinates, 
i.e. a negative value of $\epsilon\approx-1$ in the simple $\epsilon$-model. 

Although the parameterization of clustering evolution by means of the $\epsilon$-model is useful for 
{\it characterizing} the measurements as a function of redshift, 
it does not provide good physical insight into evolution governed by the clustering of 
dark matter halos \citep[see][]{giavalisco98,mccracken01}. Galaxy clustering evolution can be described more precisely 
by
\begin{equation}
\label{eq:xi}
\xi_{gal}(z,r)=D^2(z)b^2(z)\xi_m(0,r),
\end{equation}
where $D(z)$ is the linear cosmological growth rate \citep[see][]{carroll92}, $b(z)$ the evolution of the bias, 
and $\xi_m(0,r)$ the correlation function of the underlying matter distribution at $z=0$. 
Since $b(z)$ is related to the nature of the mechanism through which the galaxies were formed, measurements of 
$\xi_{gal}(z,r)$ can be used to constrain structure formation models. 

In the galaxy conservation model, objects are formed by means of monolithic collapse 
at arbitrarily high redshift, and their clustering evolution is described solely by the cosmological growth of density 
perturbations \citep{fry96}. In this model, bias evolves as 
\begin{equation}
b(z)=1+(b_0-1)/D(z), 
\end{equation}
where $b_0\equiv(\sigma_{8,gal}/\sigma_{8,m})$ and $\sigma_8$ is the rms fluctuation amplitude inside a sphere of 8 $h^{-1}$ Mpc 
radius. Taking $r_{0,m}(0)=5$ $h^{-1}$ Mpc for the present-day correlation length of the dark matter 
from the GIF/VIRGO $N$-body simulations of \citet{jenkins98} (thick solid line in Fig. \ref{fig:zoo}) and $\bar{r}_{0,gal}(0)=8.7$ $h^{-1}$ Mpc for ellipticals, we find $\sigma_{8,m}=0.9$ and $\sigma_{8,gal}=1.5$ corresponding to 
$b_0\approx1.65$. This model is indicated in Fig. \ref{fig:zoo} (thin solid line). Analogous to the above arguments against 
simple stable or linear clustering, extrapolating the clustering of local ellipticals to $z\sim1$ in the galaxy conservation model 
does not fit the observed extreme clustering of EROs and powerful radio galaxies. On the other hand, this scenario shows good agreement with the 
$r_0\sim6$ $h^{-1}$ Mpc measured for lower luminosity radio sources and QSOs at $z\sim1$.  

Crucial to the picture that is developing may be the recent results of \citet{wilson02}, who studied the clustering 
of $(V-I)$-selected $L_*$ early-type galaxies in the redshift range $0.2<z<0.9$. This author found that these galaxies cluster slightly more strongly compared to the field, with a best-fitting $\epsilon$-model of $\epsilon=0$ and $r_0=5.25\pm0.28$ $h^{-1}$ Mpc. This is in agreement with the correlation length of local $L_*$ early-types in the 2dFGRS. \citet{wilson02} remarks 
that this measurement is inconstent with the large $r_0$ found for EROs, which are also believed to be $L\sim L_*$ early-type galaxies. 
The value of $r_0$ for EROs and radio galaxies could be spuriously high due to uncertainties in their redshift distributions which is 
not included in the quoted errors, although the selection functions of both EROs and powerful radio galaxies are considered to be 
understood relatively well \citep[e.g.][]{dp90,daddi01,mc01}. Alternatively, EROs and radio galaxies at $z\sim1$ may be much more strongly clustered because they correspond to a population of massive, bright cluster galaxies in the process of formation.
If FRII radio galaxies and EROs are indeed the distant analogs of local $L\sim L_*$ early-types, 
they are becoming considerably more biased tracers of the underlying galaxy distribution with redshift, 
while this galaxy distribution itself probably traces the dark matter distribution with relatively constant bias.  
Interestingly, (semi-) analytic models and $N$-body simulations are able to explain this 
bias evolution and the large inferred $r_0$ at $z\sim1$ of massive ellipticals, if the assumption that galaxies are conserved 
quantities (i.e. closed-box systems) is relaxed. These hierarchical merging models 
\citep[e.g.][and references therein]{mo96,matarrese97,moscardini98,kauffmann99b,moustakas02,mo02} 
prescribe that for certain types of objects bias can grow stronger with redshift than the growth of 
perturbations, resulting in a $r_0$ that is constant or even increasing with redshift. 

In the (transient) model of \citet{matarrese97} it is assumed that the mass of the dark matter halo also determines the 
physical parameters of the galaxy that it contains. Based on the work of \citet{mo96} and the formalism of \citet{press74}, 
\citet{matarrese97} derive that the bias in such a model evolves as 
\begin{equation}
b(z)=1-1/\delta_c+[b_0-(1-1/\delta_c)]/D(z)^\beta,
\end{equation}
where $\delta_c=1.686$ is the critical linear overdensity for spherical collapse \citep[but see also][]{lilje92}. The parameters  
$b_0$ and $\beta$ depend on the minimum mass of the halo, and we have used the COBE-normalized ($\Lambda$CDM) values for $\mathcal{M_{\mathrm{min}}}=10^{12-14}M_\odot$ given by \citet{moscardini98} to plot this model in Fig. \ref{fig:zoo} (thin dashed lines). 
We find that the $\mathcal{M_{\mathrm{min}}}=10^{14}M_\odot$ model is able to fit the measurements at both $z\sim1$ and $z=0$.  
Likewise, the model with $\mathcal{M_{\mathrm{min}}}=10^{13}M_\odot$ has been found to fit the spatial clustering of QSOs 
relatively well \citep{croom01}, although several serious caveats exist \citep[see][]{reng98,croom01}. Most importantly, 
the assumption that there always exists a simple relationship between the mass of the dark matter halo and the property by 
which a galaxy is selected may not be valid. 

In Fig. \ref{fig:zoo} we have also indicated the predicted evolution of the clustering of early-type 
galaxies (thick dashed line) from the $\Lambda$CDM-models of \citet{kauffmann99a,kauffmann99b}
\citep[see also][]{somerville01}, normalised to $r_0$ found for local ellipticals. 
An important feature of the models presented in \citet{kauffmann99b} is that one naturally expects a 
dip in $r_0$ between $z=0$ and $z\approx1$, if 
structure is probed by galaxies of intermediate luminosities residing in haloes of masses 
$10^{11-12}M_\odot$ that have formed early and are unbiased tracers of the overall mass distribution.         
However, these simulations also show that this dip is very sensitive to sample selection criteria: massive early-type 
galaxies exhibit no dip in clustering between $z=0$ and $z\approx1$, because they occur in rare, very masssive 
haloes of $10^{13-14}M_\odot$ which are strongly biased locally, and which become even stronger 
biased with redshift. The agreement of this model with the results presented in this paper and the 
results of \citet{daddi01} and \citet{mc01} is striking. Although promising, some discrepancies between the model and 
the observations remain. 
For instance, \citet{daddi01,daddi02} find strong disagreement between the model  
and the high observed space density of EROs, seemingly consistent with the purely passive evolution of 
local ellipticals. Furthermore, current merging models generally predict that these galaxies should have 
experienced recent star-formation activity, while this is not observed. It may become possible to still reconcile the 
observations with the $\Lambda$CDM merging models if, for example, the merging is accompanied by little star-formation \citep{daddi01}. 
Also, the EROs are found to have relatively old stellar populations of $\gtrsim3$ Gyr that show no indications  
of recent formation processes. However, \citet{moustakas02} point out that the relatively old ages of their stellar populations   
do not automatically imply similar ages for the host galaxies.    

Despite the success of current hierarchical models in predicting the evolution of bias for these massive galaxies, we would like to 
point out that galaxy conservation or linear/stable clustering evolution could still be able to explain the measurements if 
EROs and/or powerful radio galaxies are solely found in rich Abell-type clusters with (present-day) $r_0\sim15-25$ $h^{-1}$ Mpc. 
As we have shown there is substantial evidence that this may be the case for, at least, the powerful radio galaxies, and future 
data may show whether this also holds for (a subset of) the population of EROs. 
 
At the highest redshifts, clustering of LBGs at $3\lesssim z\lesssim4$ indicate that these objects can be connected to local   
ellipticals in a galaxy conservation scenario. However, it is now believed that LBGs 
probably occupy much less massive halos of $10^{11-12}M_\odot$ than those that contain  
local massive galaxies, suggesting that if these objects are to be the progenitors of local ellipticals, they must have 
accumulated a considerable amount of mass \citep{adelberger00,moustakas02}. 

\subsubsection{Clustering and the occurrence of AGN at high $z$}

Fig. \ref{fig:zoo} suggests that the clustering evolution of {\it active} galaxies in general 
is considerably different from that of ordinary galaxies. Albeit at a lower amplitude, the clustering of QSOs 
also shows a trend of constant or slightly increasing amplitude with redshift, very similar to the trend that we  
derive for the clustering of the most massive ellipticals.  
According to the standard paradigm, AGN are powered by the accretion of matter onto a 
(super-)massive blackhole \citep[e.g.][]{rees84}. This fuelling mechanism may very well be 
associated with the injection and accretion of gas during major merging events, and thus, the 
occurrence of AGN seems to be logically linked to the hierarchical scenarios for structure formation.
Recently, in a series of papers \citep{haehnelt98,hk00,kh00,kh02} the simulations of \citet{kauffmann99b} were 
extended to a unified model for the evolution of both galaxies and quasars.
In their model, elliptical galaxies, supermassive black holes and starbursts are formed during major merging events, 
in which a fraction of the available gas is used to trigger quasar activity by accretion for about $10^7$ years, 
and the remaining gas is converted into stars in a single short burst. This model succesfully reproduces 
the evolution of cold gas that is derived from observations of damped Ly$\alpha$ systems, the 
luminosity functions and clustering properties of QSOs from the 2dF QSO survey, and the relation between 
bulge velocity dispersion and black hole mass that has been found in demographic studies of black holes in 
nearby galaxies \citep[e.g.][]{kormendy95,magorrian98,gebhardt00}. 

Although it has yet remained unknown exactly what processes cause the physical differences between radio-quiet and 
radio-loud AGN, recent results indicate that the hosts of all powerful AGN (both radio-loud and radio-quiet) 
are almost exclusively $L\gtrsim L_*$ ellipticals \citep[see][and references therein]{dunlop02}. 
However, the same studies also indicate that while radio-quiet AGN hosts can have black holes with masses of 
$10^{6-10}M_\odot$, the radio-loud sources are cleanly confined to black hole masses $M_{bh}\gtrsim5\times10^8M_\odot$. 
Furthermore, in the regime of extreme radio luminosities that lie well beyond the FRI/FRII luminosity-break, the power 
needed can only be achieved by blackholes with $M_{bh}>10^9M_\odot$, requiring host masses of $>10^{12}M_\odot$ that imply $L>L_*$ 
luminosities \citep{dunlop02}. 
This may explain why the most powerful NVSS sources are extremely clustered compared to the, on average, less massive hosts of QSOs. 
This is supported by the fact that the radio sources in our lower radio luminosity bin have a correlation length similar to that of 
QSOs at $z\sim1$, while both populations are still clustered more strongly compared to the field at $z\sim1$. 
We conclude that the masses of the haloes, host galaxies, and black holes that are probed by the most powerful radio sources 
are among the most massive objects in the Universe, possibly formed through massive mergers in hierarchical fashion.

\section{Summary}

The main conclusions that can be drawn from our analysis are the following:\\
$\bullet$ Below $\sim6\arcmin$ \wt is dominated by the size distribution of multi-component radio sources. 
A simple model of the physical size distribution of FRII radio galaxies is able to 
explain the observed enhancement of the cosmological clustering signal.\\
$\bullet$ The amplitude of the angular two-point correlation function of radio 
sources increases with increasing radio flux, corresponding to a similar increase in $r_0$ with increasing average radio power of the 
samples. This suggests that powerful FRII radio galaxies are intrinsically more strongly 
clustered than the average population of radio galaxies at $z\sim1$. This is consistent with the extremely rich environments in which high redshift  
FRIIs are generally found.\\
$\bullet$ The correlation lengths of powerful radio galaxies and EROs are of comparable magnitude  
and both are associated with massive ellipticals at $z\sim1$. This suggests that we could be looking at  
identical objects at different stages of their evolution, implying that AGN activity is an important phase 
in the evolution of massive galaxies in general.\\
$\bullet$ The evolution that we infer for the clustering of massive ellipticals between $z\sim1$ and $z\sim0$ is in 
agreement with predictions from hierarchical models for structure formation, because they can account for the 
observed lack of evolution in $r_0$. However, the large correlation length of powerful radio galaxies at $z\sim1$ is also consistent 
with galaxy conservation models if they are primarily associated with rich, Abell-type clusters.\\ 

\begin{acknowledgements}
We would like to thank Chris Blake, Emanuele Daddi, Matt Jarvis, Melanie Johnston-Hollitt and Jaron Kurk for productive discussions   
and reading through the text. We also thank the referee for very helpful comments.   
\end{acknowledgements}

\bibliographystyle{aa}
\bibliography{overzier}

\end{document}